\documentclass[12pt]{article}

\usepackage{xcolor}
\usepackage[normalem]{ulem}

\usepackage{mathrsfs}
\usepackage{epsfig, graphicx,subfigure,hhline}
\usepackage{epstopdf} 
\usepackage{color,soul,xcolor,enumerate}
\usepackage{url}
\usepackage{multirow, filecontents}
\usepackage{array}
\usepackage{calc,mathtools}
\usepackage{float}
\usepackage{cancel}
\usepackage[normalem]{ulem}  
\usepackage{a4wide}
\usepackage[colorlinks=true,linkcolor=blue,citecolor=red]{hyperref}
\usepackage{amssymb,amsthm,empheq,bbold}
\usepackage[ruled,vlined,linesnumbered]{algorithm2e}
\usepackage{caption}

\numberwithin{equation}{section}

\theoremstyle{plain}

\allowdisplaybreaks

\newcommand{\be}{\begin{equation}}
	\newcommand{\ee}{\end{equation}}

\newcommand{\pa}{\partial}

\newcommand{\bx}{\mathbf{x}}
\newcommand{\bk}{\mathbf{k}}

\newcommand{\nn}{\nonumber}

\begin{document}
	
	\title{Turing instability and 2-D pattern formation in reaction-diffusion systems derived from kinetic theory}

	\author{Stefano Boccelli$^{1}$, Giorgio Martalò$^{2}$, Romina Travaglini$^{3,4*}$\\[1em]
		{$^1${\footnotesize Planetary Magnetospheres Laboratory, NASA Goddard Space Flight Center}}\\{\footnotesize 8800 Greenbelt rd, Greenbelt, MD 20771, USA}
		\\{\footnotesize stefano.boccelli@nasa.gov}\\[0.5em]
		$^2${\footnotesize Dipartimento di Matematica "Felice Casorati", Università di  Pavia}\\{\footnotesize Via Ferrata 5, 27100, Pavia, Italy}
		\\{\footnotesize giorgio.martalo@unipv.it}\\[0.5em]
		$^3${\footnotesize INDAM -- Istituto Nazionale di Alta Matematica "Francesco Severi"}\\{\footnotesize Piazzale Aldo Moro 5, 00185, Roma, Italy}
		\\
		[0.5em]
		$^4${\footnotesize Dipartimento di Scienze Matematiche, Fisiche e Informatiche, Università di  Parma}\\{\footnotesize Parco Area delle Scienze 53/A, I-43124, Parma, Italy}
		\\{\footnotesize romina.travaglini@unipr.it}
		\\{\footnotesize * corresponding author}
	}
	
	\date{}
	\maketitle

	\begin{abstract}
		We investigate Turing instability and pattern formation in two-dimensional domains for two reaction-diffusion models, obtained as diffusive limits of kinetic equations for mixtures of monatomic and polyatomic gases. The first model is of Brusselator type, which, compared with the classical formulation, presents an additional parameter  whose role in stability and pattern formation is discussed. In the second framework, the system exhibits standard nonlinear diffusion terms typical of predator–prey models, but differs in reactive terms. In both cases, the kinetic-based approach proves effective in relating macroscopic parameters, often set empirically, to microscopic interaction mechanisms, thereby rigorously identifying admissible parameter ranges for the physical description. Furthermore, weakly nonlinear analysis and numerical simulations extend previously known one-dimensional results and reveal a wider scenario of spatial structures, including spots, stripes, and hexagonal arrays, that better reflect the richness observed in real-world systems.
	\end{abstract}
	
	\smallskip
	
	\noindent{\bf Keywords:} Reaction-diffusion equations, Kinetic theory, Weakly Nonlinear Analysis, Pattern formation, 
	
	\smallskip
	
	\noindent{\bf MSC Classification:} 35K57, 35Q92, 82C40

	\maketitle

	\section{Introduction}
	
	Reaction-diffusion equations play a crucial role in various areas of applied sciences, offering a powerful framework for modeling diverse {  spatio-temporal dynamics} in complex systems, that involve both interactions among the components and diffusive phenomena. 
	{  Since the pioneering work of Alan Turing in 1952 \cite{turing1990chemical}, who first demonstrated how diffusion could destabilize a uniform equilibrium and lead to the emergence of stationary patterns, many reaction–diffusion models have been formulated at the macroscopic level by specifying partial differential equations for the concentrations or densities of interacting species. 
		\\
		Diffusion coefficients and reaction rates are introduced or fitted through phenomenological considerations. A well-known example is the Brusselator model, a classical autocatalytic reaction–diffusion system developed by Prigogine and collaborators \cite{lefever1988brusselator,prigogine1968symmetry}. It has been extensively studied for its capacity to generate stable spatial patterns and to exhibit Turing instability under suitable parameter regimes \cite{gambino2013turing,pena2001stability}.
		Moving forward, these models have been applied to a wide range of phenomena across disciplines. Examples include morphogenesis in autocatalytic chemical reactions \cite{gambino2013turing,pena2001stability}, vegetation patterns in arid landscapes \cite{abbas2025pde,consolo2025vegetation}, epidemic spread \cite{banerjee2023spatio,della2025spatiotemporal}, lesion formation in autoimmune dynamics \cite{bisi2024chemotaxis, gargano2024cytokine}, and even socioeconomic processes such as market dynamics and migration \cite{banerjee2021nonlocal}.\\
		Beyond their broad applicability, classical reaction–diffusion systems possess a well developed mathematical theory explaining how patterns emerge from uniform states. In the two-species activator–inhibitor framework introduced by Turing \cite{turing1990chemical} and extensively applied to biological systems in \cite{murray2003spatial}, diffusion-driven instability occurs when a homogeneous steady state, stable in the absence of diffusion, becomes unstable once diffusion is included, typically requiring a strong disparity between diffusion coefficients. Near the instability threshold, the dominant spatial modes select characteristic wavelengths, and in two-dimensional domains these modes can interact to produce a variety of stationary structures, including stripes, spots, labyrinthine patterns, and hexagonal arrays, as demonstrated numerically in early works like \cite{muratov1996scenarios} and experimentally in chemical systems (e.g., \cite{yang2006turing}).\\
		While one-dimensional settings allow only simple periodic configurations, two dimensions enable mode resonances, symmetry-breaking interactions, and the coexistence or competition of multiple pattern analyzed in depth through bifurcation and weakly nonlinear theory \cite{judd2000simple}. Weakly nonlinear (amplitude-equation) analysis has been a central tool for understanding this pattern-selection mechanism, for predicting the stability of different morphologies, and for developing more systematic criteria for spot–stripe selection and related transitions, as explored for instance in \cite{marquez2014selection}.
		\\
		These classical approaches have proven highly successful in reproducing observed patterns across many domains, nevertheless, they often do not account for the underlying microscopic interactions between individual particles, agents, or molecules. As a result, the macroscopic parameters, such as diffusion coefficients and reaction constants, may lack a clear mechanistic interpretation. This limits the model’s predictive power, particularly in contexts where one seeks to relate observed macroscopic phenomena to their underlying physical or chemical processes.\\
		In contrast, one can formulate macroscopic reaction–diffusion models directly from a microscopic description. This strategy allows the fundamental microscopic phenomena of interest to be modeled explicitly, with the macroscopic equations arising only afterward as collective effects of the microscopic dynamics. In doing so, the need to assign empirical parameter values is reduced or eliminated; instead, the resulting macroscopic expressions are grounded in the underlying microscopic physics of the problem.\\
		To this aim, extended kinetic theory offers a powerful multiscale methodology, which has been employed in a variety of physical, chemical, and biological problems through the years (see for example \cite{bellomo2022towards,cercignani1988boltzmann, mascali2017exploitation, struchtrup2022twenty}).
	By starting from a mesoscopic description of particle interactions, it is possible to rigorously derive macroscopic reaction–diffusion systems through suitable hydrodynamic limits. These tools have been widely used in different frameworks, from classical Boltzmann theory of gas dynamics \cite{bisi2006reactive,bisi2022reaction,lachowicz2002microscopic} to the models of cells and tissues (see e.g. \cite{burini2019multiscale} and references therein). Similarly, the micro-macro connection has been investigated in the case when the reactive collision operators are phenomenological models, such as Fokker-Planck or BGK \cite{spigler1992reaction,spigler1993asymptotic,spigler1994bgk}, or for discrete velocity models \cite{zanette1993linear}.\\
	This approach not only provides explicit expressions for macroscopic parameters in terms of microscopic quantities but also reveals how collective behavior emerges from individual interactions.}
	For the sake of completeness, it is important to { remark} that, from a broader perspective, kinetic theory offers the possibility of working across multiple scales, { ultimately} leading to
	a deeper understanding of collective phenomena and allowing, for example, insight into
	how a population is distributed around the { mean value. Some} steps in this direction have
	recently been taken in  \cite{bondesan2025lotka,martalo2025individual}.\\
	{  In this paper, we focus on two such reaction-diffusion systems \cite{bisi2022reaction,martalo2024reaction}, recently derived from kinetic models of gas mixtures involving monatomic and polyatomic species in proper multiple scale regimes.  These models consider a mixture of interacting gases, initially modeled at the microscopic level using kinetic equations. From the microscopic formulation, the authors subsequently obtain a macroscopic formulation in the form of a reaction-diffusion system of equations. Unlike classical reaction–diffusion systems where diffusion and reaction rates are free parameters, the models derived in \cite{bisi2022reaction,martalo2024reaction} inherit nonlinear and cross-diffusion structures that arise directly from collision dynamics at the microscopic scale, leading to constraints and couplings absent in phenomenological models.\\
		Specifically, these} references show how  linear and nonlinear diffusion terms can be obtained through a rigorous limit of Boltzmann-like kinetic equations. More precisely, in the continuum limit, in \cite{bisi2022reaction} components densities are governed by a Brusselator-type description, where a further parameter appears in addition to those present in the classic model for autocatalytic reactions; whereas, in \cite{martalo2024reaction} a model with nonlinear cross diffusion is derived, reproducing the diffusive terms of \cite{conforto2018reaction} but differing from it in the reactive contributions. In both cases, the parameters in the macroscopic setting can be expressed in terms of the microscopic quantities, such as mass, internal energy and collision frequency{ , allowing for a physically grounded parameter selection and a deeper understanding of pattern formation mechanisms.}\\
	In the study of macroscopic reaction-diffusion systems, considerable attention is devoted to Turing instability, which represents a key mechanism for the emergence of spatio-temporal patterns. Specifically, Turing instability refers to the process by which a stable stationary solution becomes destabilized in the presence of diffusion, leading to the formation of stable non-uniform structures. This issue has been addressed in both \cite{bisi2022reaction,martalo2024reaction}, where macroscopic coefficients are explicitly linked to microscopic parameters, showing that system properties such as stability and the onset of Turing instability depend on the internal energy levels of monoatomic and polyatomic species. In those works, both the analysis and the numerical investigation were carried out in one-dimensional domains. 
	\\
	Moving from one-dimensional analysis to a two-dimensional framework is essential, as it enables the emergence of richer and more complex spatial patterns -- such as spots, stripes, and hexagonal arrays -- that cannot appear in one dimension. Moreover, extending to two-dimensional models brings simulations closer to the spatial complexity observed in real-world systems.
	\\
	The aim of this paper is twofold: first, to revisit the derivation from kinetic equations in order to highlight how the dependence on microscopic dynamics naturally imposes constraints on the choice of parameters; second, to analyze the phenomenon of Turing instability and pattern formation in two-dimensional domains for varying parameters within the validity range.
	The latter objective involves the study of interactions between multiple modes and wavevectors, which give rise to a more intricate landscape of possible instabilities and transitions.
	\\
	This can be addressed through a weakly nonlinear analysis \cite{Ouyang2000}, which allows us to predict the emergence and stability of different spatial configurations. Such an approach, which is widely used in describing the formation of two dimensional patterns in various fields ad cellular biology \cite{bisi2025derivation}, population dynamics \cite{li2018pattern} or vegetation-water interplay \cite{consolo2025vegetation}, has been extensively explored for Brusselator-type models \cite{gambino2013turing,pena2001stability,verdasca1992reentrant} in the context of chemical reactions. Here, we extend this investigation by considering, on the one hand, the model for mono/polyatomic gas mixtures, which introduces an additional coefficient, and on the other hand, by applying the same methodology, the model in \cite{martalo2024reaction}.
	\\
	In both cases, as in the basic Turing instability analysis, we are able to derive conditions on the microscopic parameters that give rise to a variety of two-dimensional scenarios.
	\\
	The paper is organized as follows: we briefly recall the derivation of the two models as diffusive limit of proper kinetic equations in Section \ref{derivation}. The Turing instability and the pattern formation for the Brusselator-type model and the description with cross diffusion are { discussed} in Section \ref{brusselator_analysis} and \ref{cross_analysis}, respectively: the emergence of non-uniform stable solutions is investigated through weakly nonlinear analysis, and confirmed and extended  numerically. Some concluding remarks are given in Section \ref{conclusions}.
	
	\section{Kinetic equations and diffusive limits}
	\label{derivation}
	
	We recall here the physical setting used in \cite{bisi2022reaction, martalo2024reaction} to derive some reaction-diffusion systems as diffusive limit of kinetic equations in gas dynamics.\\
	{  We first provide some preliminaries on the physical modeling adopted here to describe interacting mixtures of polyatomic species \cite{giovangigli2012multicomponent,groppi1999kinetic}. Each gas species, say $C^i$, may have a number $l^i$ of possible discrete values for the internal energy. We denote each one of these values by $E^i_j$, $j=1,\ldots,l^i$. As a consequence, we treat the $i$-th species as composed by $l^i$ components $C^i_j$, $j=1,\ldots,l^i$, being the component $C^i_j$ characterized by the internal energy $E^i_j$.}
 	{ We report here a general encounter expressed as follows
 	\begin{equation} C^i_j+C^h_k\leftrightarrows C^l_m+C^n_p.\label{C0CollGen}\end{equation}
 	In the case of mechanical collisions, in which molecules do not change their  nature, we may have elastic scattering, when there is not energy dissipation due to absorption by the particles, and in the formulation \eqref{C0CollGen} we have $(i,j)=(l,m)$ and $(h,k)=(n,p)$. If, instead, there is a variation of the internal energies of the pair of impinging particles, we have the collision \eqref{C0CollGen} with $i=l$ and $h=n$, but $j \not= m$ and/or $k \not= p$. Finally, in case of chemical encounters, there is also a mass transfer and particles change the gas species they belong to. When this happens we have collision \eqref{C0CollGen} with $(i,h)\neq(l,n)$ and $i\neq n$. \\		
 	Indicating by $f_j^i(t,\mathbf{x},\mathbf{v})$ the distribution function associated to the component $C_j^i$ depending on time $t$, position $\mathbf{x}$, and molecular velocity $\mathbf{v}$, the Boltzmann-type operator for encounter \eqref{C0CollGen}  takes the following form (dependence on time and space is omitted)
 	\begin{equation}
 		\begin{aligned}
 			\mathcal{Q}_i^{XX}(f_j^i,f_k^h,f_m^l,f_p^n)=\int_{\mathbb{R}^3}\int_{\mathbb{S}^2}&\mathcal{H}(g^2-\delta_{ih}^{ln})g\sigma(g,\hat{\boldsymbol{\Omega}}\cdot\hat{\boldsymbol{\Omega}}^\prime)\\
 			&\times\left[\left(\dfrac{m_im_h}{m_lm_n}\right)^3f_m^l(\mathbf{v}^\prime)f_p^n(\mathbf{w}^\prime)-f_j^i(\mathbf{v})f_k^h(\mathbf{w})\right]d\mathbf{w}d\hat{\boldsymbol{\Omega}}^\prime\,,
 		\end{aligned}
 	\end{equation}
 	where
 	\begin{itemize}
 		\item[-] $(\mathbf{v},\mathbf{w})$ are the ingoing microscopic velocities of particles with mass $m_i$ and $m_h$, respectively;
 		\item[-]  $(\mathbf{v}^\prime,\mathbf{w}^\prime)$ are the post interaction microscopic velocities of particles having mass $m_l$ and $m_n$ ($l=i$ and $n=h$ in mechanical encounters);
 		\item[-] the heaviside function $\mathcal{H}$ depends on the modulus of relative velocity $\mathbf{g}=\mathbf{v}-\mathbf{w}$ and on the energy gap $\delta_{ih}^{ln}=2m_i\,m_h\,(E_m^l+E_p^n-E_j^i-E_k^h)/(m_i+m_h)$ (the gap is null in elastic collisions);
 		\item[-] the cross sections $\sigma$ depends on the modulus of the ingoing relative velocity and on the deflection angle $\hat{\boldsymbol{\Omega}}\cdot\hat{\boldsymbol{\Omega}}^\prime$, being $\hat{\boldsymbol{\Omega}}$ the direction of vector $\mathbf{g}$.
 \end{itemize}}
	{ The gaseous mixture considered in in \cite{bisi2022reaction, martalo2024reaction} is composed by two constituents} diffusing in a host medium. The first constituent $Y$ is assumed to be polyatomic with two different energy levels $E_1$ and $E_2$; therefore, it can be split in two components $Y_1$ and $Y_2$. The second gas $Z$ is supposed to be monatomic, endowed with a unique energy level $E_Z$. We indicate by $m_Y$ and $m_Z$ the masses of polyatomic and monatomic gas particles, respectively.\\
	The background medium is a mixture whose gases are much denser than the previous ones. We suppose that this mixture is composed by three components $A$, $B$ and $C$, with masses $m_A$, $m_B$ and $m_C$. We assume that their distributions are maxwellian functions with unit temperature and zero mean velocity
	\begin{equation}
		f_i = n_i\mathcal{M}_i(\mathbf{v})\,,\qquad\mathcal{M}_i(\mathbf{v})=n_i\left(\dfrac{m_i}{2\pi}\right)^{3/2}\exp\left(-\dfrac{m_i|\mathbf{v}|^2}{2}\right)\,,
	\end{equation}
	$i=A,B,C$; the number densities $n_i$ are supposed to be constant.
	
	The evolution of the distribution functions $f_1$, $f_2$ and $f_Z$ of constituents $Y_1$, $Y_2$ and $Y_Z$ is governed by Boltzmann-like equations, where the collision operator takes account of elastic, inelastic and reactive interactions. More precisely, in addition to elastic collisions, the following transitions occur
	\begin{eqnarray}
		A+Y_1&\rightarrow &A+Y_2\\
		Z+Y_2&\rightarrow &Z+Y_1\\
		B+Y_1&\rightleftharpoons &A+C\label{reac1}\\
		Y_1+Y_1&\rightleftharpoons &Z+B\,.\label{reac2}
	\end{eqnarray}
	
	The kinetic equation for $f_i$ ($i=1,2,Z$) reads as
	\begin{equation}
		\dfrac{\partial f_i}{\partial t}+\mathbf{v}\cdot\nabla_\mathbf{x}f_i = \sum_{j=A,B,C}\mathcal{Q}_i^{EL}(f_i,n_j\mathcal{M}_j)+\sum_{j=1,2,Z}\mathcal{Q}_i^{EL}(f_i,f_j)+\mathcal{Q}_i^{IN}(\mathbf{f})+\mathcal{Q}_i^{CH}(\mathbf{f})\,,
	\end{equation}
	where $\mathbf{f}$ is the vector collecting all the distribution functions. The first sum models the elastic interactions with the host medium, the second one accounts for elastic interactions between components of the mixture. $\mathcal{Q}_i^{IN}$ and $\mathcal{Q}_i^{CH}$ describe the contributions of inelastic transitions and chemical reactions, respectively.\\
	We refer to \cite{bisi2022reaction,martalo2024reaction} for a detailed description of the Boltzmann-type operators; here, we recall only that the Maxwell molecule intermolecular potential is adopted, leading to constant collision frequencies.
	
	\subsection{Linear diffusion}
	
	Starting from the kinetic description, it is possible to derive suitable reaction-diffusion systems for the number densities, performing a proper hydrodynamic limit. To this aim, we introduce a scaling in terms of a small parameter $\varepsilon$ (standing for the Knudsen number), assuming a multi-scale interaction process.\\
	This approach can be used to derive a reaction-diffusion system similar to the Brusselator one; in particular, we assume that
	\begin{itemize}
		\item [-] the dominant phenomenon is constituted by elastic collisions with the host medium. Collisions of the two components of $Y$ with the background are assumed of order $1/\varepsilon$; interactions involving the gas $Z$ are supposed to be faster of order $1/\varepsilon^2$;
		\item [-] the other elastic collisions provide a phenomenon of order $\varepsilon^p$ ($p\ge 0$);
		\item [-] inelastic interactions and the chemical reaction \eqref{reac1} are the slow process of order $\varepsilon$; reaction \eqref{reac2} is assumed to be faster of order 1.
	\end{itemize}
	In addition, since we focus on the effects of inelastic and chemical encounters, that are of order $\varepsilon$, we  re-scale the time by a factor $\varepsilon$, that, consequently, appears in front of the temporal derivatives.
	
	We use the same procedure proposed in previous papers and we consider a proper expansion of the distribution functions (now indicated by $f_i^\varepsilon$, $i=1,2,Z$ to highlight the dependence on the small parameter $\varepsilon$). We observe that the evolution is dominated by collisions with the background; then
	\begin{equation}
		\sum_{j=A,B,C}\mathcal{Q}_i^{EL}(f_i^\varepsilon,n_j^\varepsilon\mathcal{M}_j)=O(\varepsilon)\,,\, i=1,2\,,\qquad \sum_{j=A,B,C}\mathcal{Q}_i^{EL}(f_i^\varepsilon,n_j^\varepsilon\mathcal{M}_j)=O(\varepsilon^2)\,,\, i=Z\,,
	\end{equation}
	and hence
	\begin{eqnarray}
		f_i^\varepsilon(t,\mathbf{x},\mathbf{v})=n_i^\varepsilon(t,\mathbf{x})\mathcal{M}_i(\mathbf{v})+\varepsilon h_i^\varepsilon(t,\mathbf{x},\mathbf{v})\,,\,i=1,2\label{exp1}\\
		f_i^\varepsilon(t,\mathbf{x},\mathbf{v})=n_i^\varepsilon(t,\mathbf{x})\mathcal{M}_i(\mathbf{v})+\varepsilon^2 h_i^\varepsilon(t,\mathbf{x},\mathbf{v})\,,\,i=Z\,,\label{exp2}
	\end{eqnarray}
	thus
	\begin{equation}
		\int_{\mathbb{R}^3}h_i^\varepsilon(\mathbf{v})\,d\mathbf{v}=0,\,i=1,2,Z\,.
	\end{equation}
	By considering the weak form of the (non-dimensional) kinetic equations, by using the previous expansions \eqref{exp1} and \eqref{exp2}, and by manipulating terms of the same order, we obtain the following set of partial differential equations for number densities $n_1$ and $n_2$
	\begin{equation}
		\begin{aligned}
			&\dfrac{\partial n_1}{\partial t}-D_1\Delta_\mathbf{x}n_1=a-(b+1)n_1+dn_1^2n_2\\
			&\dfrac{\partial n_2}{\partial t}-D_2\Delta_\mathbf{x}n_2=bn_1-dn_1^2n_2
		\end{aligned}
		\label{brusselator}
	\end{equation}
	in the formal limit when $\varepsilon\rightarrow 0$.	The derivation of system \eqref{brusselator} is detailed in \cite{bisi2022reaction}.\\ The coefficients of system \eqref{brusselator} are, indeed, functions of the microscopic parameters characterizing the mixture components. More precisely,
	\begin{align}
		a &= \dfrac{e^{\Delta E_{B1}^{AC}} \left( \dfrac{m_B m_Y}{m_A m_C} \right)^{3/2} n_A n_C}{n_B}, \label{coefReaBrus} \\
		b &= \dfrac{n_A \, \Gamma\left( \dfrac{3}{2}, \Theta(\Delta E_{A1}^{A2}) \right) \nu_{A1}^{A2}}{n_B \, \Gamma\left( \dfrac{3}{2}, \Theta(\Delta E_{B1}^{AC}) \right) \nu_{B1}^{AC}}, \\
		d &= \dfrac{e^{-\Delta E_{11}^{ZB}} \, \Gamma\left( \dfrac{3}{2}, \Theta(\Delta E_{Z2}^{Z1}) \right) \nu_{Z2}^{Z1}}{ \left( \dfrac{m_Y^2}{m_B m_Z} \right)^{3/2} n_B^2 \, \Gamma\left( \dfrac{3}{2}, \Theta(\Delta E_{B1}^{AC}) \right) \nu_{B1}^{AC} },
	\end{align}
	and
	\begin{equation}
		D_1 =\Pi  \displaystyle\sum_{J \in \{A, B, C\}} \dfrac{m_J n_J \nu_{1J}}{m_Y + m_J},
		\quad
		D_2 =\Pi \displaystyle \sum_{J \in \{A, B, C\}} \dfrac{m_J n_J\nu_{2J}}{m_Y + m_J},\quad \Pi=\dfrac{\sqrt{\pi}\,\nu_{B1}^{AC} }{2 m_Y n_B \, \Gamma\left(\dfrac{3}{2}, \Theta\big(\Delta E_{B1}^{AC}\big)\right)} 
		\label{coefD1D2Bruss}
	\end{equation}
	
	where
	\begin{itemize}
		\item[-] $\nu_{IJ}$ is the collision frequency associated to an elastic collision between components $I$ and $J$;
		
		\item[-] $\nu_{IJ}^{HK}$ is the collision frequency associated to a (inelastic or chemical) transition  leading $I,J$ into $H,K$;
		
		\item[-] $\Gamma\left(\frac{3}{2}, \Theta(\Delta E)\right)$ is the incomplete Gamma function of order $3/2$, where the second argument $\Theta(\Delta E)$ is a proper threshold depending on the energy gap $\Delta E$;
		
		\item[-] $\Delta E_{IJ}^{HK}$ indicates the energy gap associated to a (inelastic or chemical) transition  from states $I,J$ to $H,K$;
		
	\end{itemize}

	It is important to underline that, through the derivation, an explicit expression of $n_Z$ as quadratic function of $n_1$ is deduced; this algebraic relation allows to reduce the number of variables in the diffusive limit.\\
	{ In the following, the stability properties will be investigated in a regular open domain; if we suppose that the distribution functions $f_i^\varepsilon$ satisfy given initial conditions at time $t=0$ and specular reflection conditions at the boundary, then, as proved in \cite{bisi2006reactive}, the macroscopic densities exactly obeys equations \eqref{brusselator}, with proper initial data and no-flux conditions at the boundary.\\
	The procedure briefly outlined here shows that the coefficients of the macroscopic observables are positive constants that can be explicitly expressed in terms of the underlying microscopic parameters, such as collision frequencies, particle masses, and discrete energy levels. In this way, the collective behavior encoded in the macroscopic equations remains directly related to the interaction mechanisms governing the particles dynamics at the microscopic scale. This correspondence not only enhances the physical meaning of the macroscopic coefficients but also introduces natural constraints on their admissible values, thereby delimiting the range of validity of the resulting continuum model.}\\
	As last comment, we notice that the system \eqref{brusselator} is of Brusselator-type; the unique difference consists in the occurrence of coefficient $d$, usually equal to 1 in literature \cite{lefever1988brusselator,lengyel1991modeling}. The role of this parameter, and its effects on the pattern formation, will be investigated in the next section.
	
	\subsection{Nonlinear cross diffusion}
	
	A similar procedure has been used recently to derive a system with nonlinear cross diffusion in the diffusive limit. In this case we consider a different multiple scale process, where
	\begin{itemize}
		\item [-] the dominant phenomenon is constituted by elastic collisions with the host medium, which are assumed of order $1/\varepsilon^2$;
		\item [-] the other elastic collisions provide a phenomenon of order $\varepsilon^p$ ($p\ge 0$);
		\item [-] inelastic interactions and the chemical reactions proceed at order $\varepsilon$ and $\varepsilon^2$, respectively.
	\end{itemize}
	We rescale the time variable by a factor $\varepsilon^2$.\\
	The distribution functions are expanded as
	\begin{equation}
		f_i^\varepsilon(t,\mathbf{x},\mathbf{v})=n_i^\varepsilon(t,\mathbf{x})\mathcal{M}_i(\mathbf{v})+\varepsilon^2 h_i^\varepsilon(t,\mathbf{x},\mathbf{v})\,,\,i=1,2,Z\,,
	\end{equation}
	since
	\begin{equation}
		\sum_{j=A,B,C}\mathcal{Q}_i^{EL}(f_i^\varepsilon,n_j^\varepsilon\mathcal{M}_j)=O(\varepsilon^2)\,,\, i=1,2,Z\,.
	\end{equation}
	In the asymptotic limit, when $\varepsilon\rightarrow 0$, we get a reaction-diffusion system
	\begin{equation}
		\begin{aligned}
			&\dfrac{\partial N}{\partial t}-\Delta_\mathbf{x}\left(\dfrac{D_1\beta n_Z+D_2}{\beta n_Z+1}\,N\right)=a-\left(\dfrac{\beta n_Z}{\beta n_Z+1}\,N\right)+cn_Z-d\left(\dfrac{\beta n_Z}{\beta n_Z+1}\,N\right)^2\\
			&\dfrac{\partial n_Z}{\partial t}-D_Z\Delta_\mathbf{x}n_Z=-cn_Z+d\left(\dfrac{\beta n_Z}{\beta n_Z+1}\,N\right)^2\,,\label{SistCros}
		\end{aligned}
	\end{equation}
	where $N=n_1+n_2$ is the total number density of the polyatomic component and the use of an algebraic relation relating $n_1$, $n_2$ and $n_Z$ has been made{ ; as shown in the previous subsection, equations are coupled with proper initial and boundary conditions}. Once again, the parameters are positive constants, depending on microscopic parameters as provided below
	\begin{equation}
		a = \dfrac{e^{\, E_{B1}^{AC}} \left(\dfrac{m_B m_Y}{m_A m_C}\right)^{3/2} n_A n_C}{n_B},\qquad	\beta = \dfrac{
			n_A \, \Gamma\left(\dfrac{3}{2}, \Theta\left(E_{A1}^{A2}\right)\right) \nu_{A1}^{A2}
		}{
			\Gamma\left(\dfrac{3}{2}, \Theta\left(E_{Z2}^{Z1}\right)\right) \nu_{Z2}^{Z1}
		},\label{coefReaCros1} 
	\end{equation}
	\begin{equation}
		c = \dfrac{
			e^{\, E_{11}^{ZB}} \left(\dfrac{m_Y^2}{m_B m_Z}\right)^{3/2} \Gamma\left(\dfrac{3}{2}, \Theta\left(E_{11}^{ZB}\right)\right) \nu_{11}^{ZB}
		}{
			\Gamma\left(\dfrac{3}{2}, \Theta\left(E_{B1}^{AC}\right)\right) \nu_{B1}^{AC}
		},\qquad 	d = \dfrac{
			\Gamma\left(\dfrac{3}{2}, \Theta\left(E_{11}^{ZB}\right)\right) \nu_{11}^{ZB}
		}{
			n_B \, \Gamma\left(\dfrac{3}{2}, \Theta\left(E_{B1}^{AC}\right)\right) \nu_{B1}^{AC}
		},
	\end{equation}
	
	\begin{equation}
		D_1 =\Pi\sum_{J \in \{A, B, C\}} \dfrac{m_J n_J \nu_{1J}}{m_Y + m_J},\qquad
		D_2 =\Pi\sum_{J \in \{A, B, C\}} \dfrac{m_J n_J \nu_{2J}}{m_Y + m_J},\label{coefReaCros2}
	\end{equation}
	
	\begin{equation}
		D_Z =\Pi\sum_{J \in \{A, B, C\}} \dfrac{m_J n_J \nu_{ZJ}}{m_Z + m_J},
	\end{equation}
	and $\Pi$ is defined as in \eqref{coefD1D2Bruss}. Further details can be found in \cite{martalo2024reaction}.\\
	The model deduced from kinetic equations in gasdynamics presents the same diffusive terms as the predator-prey model proposed in \cite{conforto2018reaction} and differs from it in the reactive contributions.
	
	\section{Analysis of the Brusselator-type model}
	\label{brusselator_analysis}
	
	We start to discuss the possible emergence of periodic solutions in the description \eqref{brusselator}, and we focus on the role of the parameter $d$ that does not appear in the classical Brusselator model.
	
	First, we consider the steady states (and their stability) in the space homogeneous case.\\
	The system admits a unique equilibrium $E=(n_1^*,n_2^*)=(a,b/ad)$; its stability is investigated through the analysis of the linearized homogeneous problem
	\begin{equation}
		\dfrac{d\, \mathbf{U}}{d\, t}=\mathbf{J}\mathbf{U}\,,
	\end{equation}
	where $\mathbf{U}=(n_1-n_1^*,n_2-n_2^*)^T$and the jacobian matrix results
	\begin{equation}
		\mathbf{J}=\left(\begin{array}{cc}
			b-1 & a^2d\\
			\\
			-b & -a^2d
		\end{array}\right)\,.
	\end{equation} 
	We notice that $\det\mathbf{J}=a^2d>0$, while the trace of $\mathbf{J}$ is negative if and only if
	\begin{equation}
		b<1+a^2\,d\,.
		\label{constraint}
	\end{equation}
	The possibility that the equilibrium becomes unstable in presence of diffusive terms will be explored under the constraint \eqref{constraint}. We can already observe how the new parameter $d$ affects both the steady state and its stability.
	
	\subsection{Turing instability}
	
\begin{figure}[ht!]
		\centering
		\includegraphics[width=0.8\linewidth]{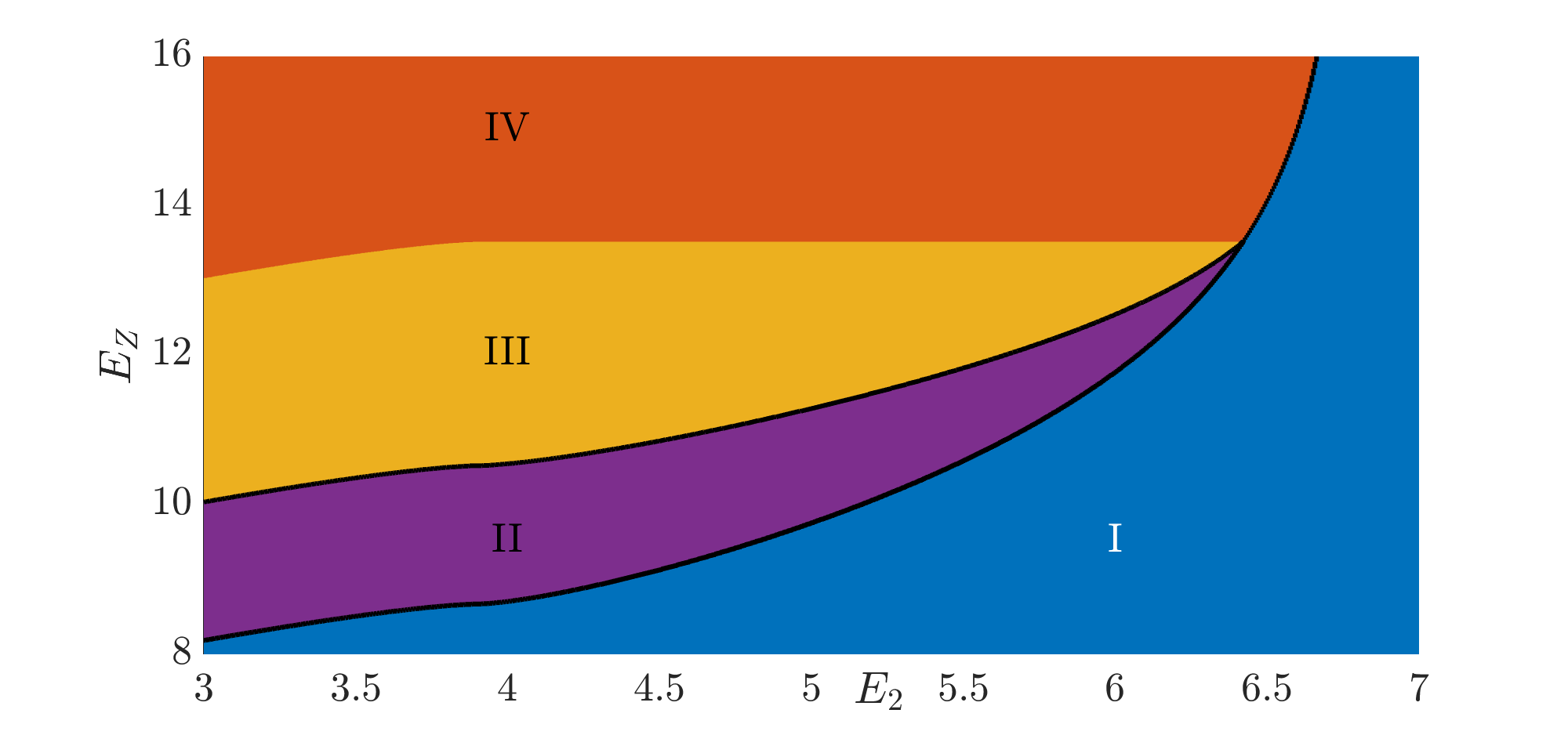}
		\caption{Values for energy levels $E_2$ and $E_Z$ relevant to Turing instability for system \eqref{brusselator}. In region I, condition \eqref{constraint} is not satisfied, in regions II, III, and IV  \eqref{constraint} holds, and either both conditions \eqref{TurBrussMac} and \eqref{TurBrussMac2}, only \eqref{TurBrussMac2}, or neither of the conditions in \eqref{TurBrussMac} and \eqref{TurBrussMac2} are satisfied. Parameters are chosen as in \eqref{ParsBrus}. }
		\label{fig:regionTurBruss}
	\end{figure}

	We consider now the system \eqref{brusselator}  in a bounded domain $\Omega$; we assume no-flux conditions at the boundary $\partial\Omega$. We have to discuss the solutions of the linearized system
	\begin{equation}
		\dfrac{\partial \mathbf{U}}{\partial t}=\mathbf{D}\Delta_\mathbf{x}\mathbf{U}+\mathbf{J}\mathbf{U}\,,\qquad\text{ in }\mathbb{R}_+\times\Omega\,,
	\end{equation}
	where the diffusion matrix results
	\begin{equation}
		\mathbf{D}=\left(\begin{array}{cc}
			D_1 & 0\\
			0 & D_2
		\end{array}\right)\,.
	\end{equation}
	We look for solutions
	\begin{equation}
		\mathbf{U}(\mathbf{x},t)=\sum_{k\in\mathbb{N}}c_ke^{\lambda_kt}\tilde{\mathbf{U}}_k(\mathbf{x})\,,
		\label{solutions}
	\end{equation}
	where the eigenfunction $\tilde{\mathbf{U}}_k$ is solution of
	\begin{equation}
		\begin{cases}
			\Delta_{\mathbf{x}}\tilde{\mathbf{U}}_k+k^2\tilde{\mathbf{U}}_k=0&\text{ in }\Omega\\
			\hat{\mathbf{n}}\cdot\nabla_\mathbf{x}\tilde{\mathbf{U}}_k=0&\text{ on }\partial\Omega\,,
		\end{cases}
		\label{eigenfunctions}
	\end{equation}
	being $\hat{\mathbf{n}}$ the outgoing unit vector at the boundary.\\
	In order to have Turing instability \cite{turing1990chemical}, the destabilization of steady state due to spatial perturbations is needed; this results in the existence of at least a wavenumber $k$, such that the corresponding $\lambda_k$	has positive real part. Each $\lambda_k$ 
	is solution of the dispersion relation
	\begin{equation}
		\lambda^2+g(k^2)\lambda+h(k^2)=0\,,
	\end{equation}
	where
	\begin{equation}
		g(k^2)=k^2\text{Tr}\mathbf{D}-\text{Tr}\mathbf{J}>0
	\end{equation}
	because of relation \eqref{constraint}, and
	\begin{equation}
		h(k^2)=\det(\mathbf{D})k^4+qk^2+\det(\mathbf{J})\,,
	\end{equation}
	being $q=(1-b)D_2+a^2dD_1$; thus, the existence of a negative solution is guaranteed by $h(k^2)<0$; this implies that we have to impose that the minimum of $h$ attained in
	\begin{equation}
		k_c^2=-\dfrac{q}{2\det(\mathbf{D})}=-\dfrac{(1-b)D_2+a^2dD_1}{D_1D_2}
		\label{minimum_x}
	\end{equation}
	must be negative, i.e.
	\begin{equation}
		h(k_c^2)=\dfrac{4\det(\mathbf{D})\det(\mathbf{J})-q^2}{4\det(\mathbf{D})}=\dfrac{4a^2dD_1D_2-[(1-b)D_2+a^2dD_1]^2}{4D_1D_2}<0\,,
	\end{equation}
	providing
	\begin{equation}
		b < -2a \sqrt{\frac{d D_{1}}{D_{2}}} + \frac{a^{2} d D_{1}}{D_{2}}+1
		\quad \text{or} \quad
		b > 2a \sqrt{\frac{d D_{1}}{D_{2}}} + \frac{a^{2} d D_{1}}{D_{2}}+1
		.
	\end{equation}
	From \eqref{minimum_x}, we have also to require that $k_c^2>0$, or equivalently $q<0$, leading to
	\begin{equation}
		b > \frac{a^{2} d D_{1}}{D_{2}}+1,
	\end{equation}
	which is compatible with constraint \eqref{constraint} if and only if $D_1<D_2$; moreover $b>1$.\\
	By summing up, the destabilization of the equilibrium occurs when 
	\begin{equation}
		b_c:=\left(1 + a \sqrt{\frac{d D_{1}}{D_{2}}}\right)^2 \;<\; b \;<\; 1 + a^{2} d\,,
		\label{TurBrussMac}
	\end{equation}
	provided that
	\begin{equation}
		d > \dfrac{4 D_{1} D_{2}}{a^{2} \left( D_{1} - D_{2} \right)^{2}}\,.
		\label{TurBrussMac2}
	\end{equation}
	{ By expressing the coefficients in terms of the microscopic parameters of the mixture through \eqref{coefReaBrus}-\eqref{coefD1D2Bruss} and, for illustrative purposes, without reference to any specific physical scenarios, assigning the following values}
	\begin{equation}\nn
		\begin{aligned}
			&m_A = 2, \quad m_B = 3.5, \quad m_C = 4, \quad m_Y= 2.5, \quad m_Z= 1.5,
		\end{aligned}
	\end{equation}
	
	\begin{equation}\nn
		\begin{aligned}
			&\nu_{A1}^{A2} = 0.008, \quad \nu_{Z2}^{Z1} = 0.3, \quad \nu_{B1}^{AC} = 0.001, \quad \nu_{11}^{ZB} = 1, \\
			&\nu_{1A} = \nu_{1B} = \nu_{1C} = \bar{\nu}_1 = 200, \\
			& \nu_{2A} = \nu_{2B} = \nu_{2C} =\bar{\nu}_2 = 15, \\
			& n_A = n_B = n_C = \bar{n}= 1.46,
		\end{aligned}
	\end{equation}
	
	\begin{equation}\label{ParsBrus}
		E_A = 4.5, \quad E_B = 3.6,\quad E_C = 4,  \quad E_1 = 3.9,
	\end{equation}
	we { identify} the values for energy levels $E_2$ and $E_Z$ leading to condition \eqref{TurBrussMac} and \eqref{TurBrussMac2} to be satisfied, as reported in Figure  \ref{fig:regionTurBruss}. In particular, we can distinguish four regions. In region I, condition \eqref{constraint} is not satisfied and the spatially homogeneous equilibrium is unstable. In regions II, III, and IV, instead, \eqref{constraint} holds, and either both conditions \eqref{TurBrussMac} and \eqref{TurBrussMac2}, only \eqref{TurBrussMac2}, or neither of the conditions in \eqref{TurBrussMac} and \eqref{TurBrussMac2} are satisfied, respectively. Consequently, pattern formation can arise only when taking parameters in region II. It should be emphasized that, within region II, the parameter $d$ is not restricted to the value 1, which once again highlights the broader spectrum of scenarios due to the presence of this additional parameter.
	
	In the following section we perform a deeper analysis of the possible spatial configurations of the model.
	
	\subsection{Weakly nonlinear analysis}
	
	Now, we intend to investigate shapes of patterning, along with their stability, compatible with system \eqref{brusselator}; this can be achieved by performing a deeper analysis of the problem, performing a high order expansion of quantities involved in the model. It is worth stressing that a weakly nonlinear analysis of the classical Brusselator problem has already been discussed in literature \cite{gambino2013turing,pena2001stability,verdasca1992reentrant}; therefore, here we focus on possible different patterns and/or modifications of their shape, due to specific choices of parameters in our kinetic-based model.
	
	We consider a Taylor expansion of system  \eqref{brusselator} up to the third
	order around the equilibrium $(n_1^*,n_2^*)$, obtaining
	\begin{equation}\label{SistComp1}
		\begin{aligned}
			\frac{\partial {\bf U}}{\partial t} &= \mathcal L\,{\bf U} + \mathcal H[{\bf U}], \quad \mbox{for} \quad
			{\bf U}=\left(\begin{array}{c}U\\V\end{array}\right)=\left(\begin{array}{c}n_1- n_1^*\\n_2-n_2^*\end{array}\right)
		\end{aligned}
	\end{equation}
	with
	\begin{equation}
		\mathcal L  ={{\mathbf D}}\Delta_{\bf x}+ {{\mathbf J}} 
	\end{equation}
	where $ {{\mathbf J}}$ and  ${{\mathbf D}}$ are the Jacobian  and the diffusion matrix, respectively, while
	\begin{equation}
		\mathcal H[{\bf U}] = 
		\begin{pmatrix}
			(-1 + b) U + \dfrac{2 b}{a} U^2 + a^2 d V + 4 a d U V+6 d U^2 V\\
			-b U - \dfrac{2 b}{a} U^2 - a^2 d V - 4 a d U V - 6 d U^2 V
		\end{pmatrix}.
	\end{equation}
	Keeping in mind the necessary conditions \eqref{TurBrussMac} for spatial pattern formation (in terms of the parameter $b$), we define the critical value allowing for Turing instability. In accordance with the choice of parameters made in \eqref{ParsBrus}, we have that the diffusion coefficient $D_1=1$, this choice streamlines the calculations without being restrictive, since conditions \eqref{TurBrussMac}-\eqref{TurBrussMac2} only depend on the ratio $D_2/D_1$; in addition, we define $\xi:=\sqrt{\frac{1}{D_2}}$, obtaining
	\begin{equation}
		b_c=\left(1 + a \xi \sqrt{d}\right)^2
	\end{equation}
	and the critical wavenumber
	\begin{equation}
		k_c^2=a \xi \sqrt{d},
	\end{equation}
	such that $\det({{\mathbf J}}-k_c^2{{\mathbf D}})=0$ and $\det({{\mathbf J}}-k{{\mathbf D}})>0$ for some wavenumber $k$, when $b>b_c$.
	
	{
		At this stage, we take advantage of the fact that the system's dynamics evolve more slowly, when the parameter $b$ is close to the critical threshold. This slower evolution allows us to study pattern formation using amplitude equations. Therefore, to better understand this scenario, we express the bifurcation parameter $b$ as follows
		\begin{equation}
			b=  b_c +\eta\,b_1+\eta^2\,b_2+\eta^3\,b_3+O(\eta^3),
		\end{equation}
		where $\eta$ is a small parameter; we expand also the solution vector ${\bf U}$  in terms of $\eta$
		\begin{equation}\label{UExp}
			{\bf U} = \eta
			\left(
			\begin{array}{c}U_1 \\ V_1\end{array}
			\right)
			+ \eta^2 
			\left(
			\begin{array}{c}U_2 \\ V_2 \end{array}
			\right)
			+ \eta^3
			\left(
			\begin{array}{c}U_3 \\ V_3 \end{array}
			\right)
			+ O(\eta^3).
		\end{equation}
		As already pointed out, when the bifurcation parameter is near its critical threshold, the amplitude of the emerging pattern evolves slowly over time. This separation of time scales enables us to distinguish between fast and slow temporal dynamics. As a result, we can introduce a multiple time scale framework, such that
		\begin{equation}\label{timeExp}
			\frac{\partial}{\partial t} = \eta \frac{\partial}{\partial T_1} + \eta^2 \frac{\partial}{\partial T_2} + O(\eta^2). 
		\end{equation}
		
		By substituting the expansions \eqref{UExp}–\eqref{timeExp} into system \eqref{SistComp1} and collecting terms of the same order in $\eta$, we obtain the following set of equations at successive orders:
		\begin{itemize}
			\item[-] order $\eta$:
			\begin{equation}\label{EqOrd1}
				\mathcal L_c\left[
				\left(
				\begin{array}{c}U_1 \\ V_1 \end{array}
				\right)\right]
				=0
			\end{equation}
			with 
			\begin{equation}
				\mathcal L_c  = \begin{bmatrix}  
					b_c-1 +  \Delta_{\bf x}& a^2d\\
					\\
					-b_c & -a^2d  +\dfrac{1}{\xi^2} \Delta_{\bf x}
				\end{bmatrix},
			\end{equation}
			\item[-] order $\eta^2$:
			\begin{equation}\label{EqOrd2}
				\frac{\pa}{\pa T_1}\left(
				\begin{array}{c}U_1 \\ V_1\end{array}
				\right)=	
				\mathcal L_c\left[
				\left(
				\begin{array}{c}U_2 \\ V_2 \end{array}
				\right)\right]
				+ \mathcal H_2\left[\left(
				\begin{array}{c}U_1 \\ V_1 \end{array}
				\right)\right]
			\end{equation}
			with 
			\begin{equation}
				\mathcal H_2\left[\left(\begin{array}{c}U_1 \\  V_1 \end{array}
				\right)\right]=\left[	\begin{array}{c}
					b_1 U_1 + \dfrac{b_{c} U_1^{2}}{a} + 2 a d U_1 V_1 + a^{2} d V_{2}
					\\[2mm]
					-b_1 U_1 - \dfrac{b_{c}}{a} U_1^{2} - 2 a d U_1 V_1  			 
				\end{array}\right]
			\end{equation}
			\item[-] order $\eta^3$:
			\begin{equation}\label{EqOrd3}
				\frac{\pa}{\pa T_1}\left(
				\begin{array}{c}U_2 \\ V_2 \end{array}
				\right)+	
				\frac{\pa}{\pa T_2}\left(
				\begin{array}{c}U_1 \\ V_1 \end{array}
				\right)=	
				\mathcal L_c \left[
				\left(
				\begin{array}{c}U_3 \\ V_3 \end{array}
				\right)\right]
				+ \mathcal H_3\left[\left(
				\begin{array}{c}U_1 \\ V_1 \end{array}
				\right),\left(
				\begin{array}{c}U_2 \\ V_2 \end{array}
				\right)\right]
			\end{equation}
			\begin{equation}
				\mathcal H_3\left[\left(
				\begin{array}{c}U_1 \\  V_1 \end{array}
				\right),\left(
				\begin{array}{c}U_2 \\V_2 \end{array}
				\right)\right]=
				\left[
				\begin{array}{c}
					b_2 U_1 + \dfrac{b_1 U_1^2}{a} + b_1 U_2 + \dfrac{2 b_c U_1 U_2}{a} 
					\\[2mm]  + d U_1^2 V_1 + 2 a d U_2 V_1 + 2 a d U_1 V_2
					\\[4mm]
					- b_2 U_1 - \dfrac{b_1 U_1^2}{a} - b_1 U_2 - \dfrac{2 b_c U_1 U_2}{a}
					\\[2mm]   - d U_1^2 V_1 - 2 a d U_2 V_1 - 2 a d U_1 V_2
				\end{array}
				\right].
			\end{equation}
		\end{itemize}
	}
	
	{
		By solving system \eqref{EqOrd1}, and leveraging the spectral properties of the operator $\mathcal{L}_c$, the solution can be expressed in terms of three dominant active pairs of eigenmodes $(\mathbf{k}_j, -\mathbf{k}_j)$, for $j = 1, 2, 3$. These modes correspond to wavevectors with equal magnitude $|\mathbf{k}_j| = k_c$ and are oriented at angles separated by $2\pi/3$, satisfying the condition $\mathbf{k}_1 + \mathbf{k}_2 + \mathbf{k}_3 = \mathbf{0}$ \cite{Ouyang2000}. The solution thus takes the form:
		
		\begin{equation}\label{Expu1}
			\left(
			\begin{array}{c}U_1\\V_1 \end{array}
			\right)=\left(
			\begin{array}{c}\rho\\1 \end{array}
			\right)\sum_{j=1}^3\left[{\mathcal W}_j(t)\,e^{i\,{\bf k_j\cdot\bx}}+\overline{\mathcal W}_j(t)\,e^{-i\,{\bf k_j\cdot\bx}}\right],
		\end{equation}
		where $\overline{\mathcal W}_j$ is the complex conjugate of $\mathcal W_j$, and
		$\rho=-\dfrac{a \sqrt{d}}{\xi + a\xi^2\, \sqrt{d} \, }
		$.
	}
	
	{
		Now, we move to discuss terms of order $\eta^2$ \eqref{EqOrd2}, that can be recast as 
		\begin{equation}\label{EqOrd2.2}
			\mathcal L_c\left[
			\left(
			\begin{array}{c}U_2 \\ V_2 \end{array}
			\right)\right]
			= 
			\frac{\pa}{\pa T_1}\left(
			\begin{array}{c}U_1 \\ V_1 \end{array}
			\right)
			- \mathcal H_2\left[\left(
			\begin{array}{c}U_1 \\ V_1 \end{array}
			\right)\right];
		\end{equation}
		the existence of a nontrivial solution \((U_2, V_2)^T\) to this non-homogeneous system \eqref{EqOrd2.2} is guaranteed by the Fredholm solvability condition. According to it, the right-hand side of equation \eqref{EqOrd2.2} must be orthogonal to the kernel of the adjoint operator \(\mathcal{L}_c^+\) of \(\mathcal{L}_c\), whose corresponding eigenvectors are given by
		
		\begin{equation}
			\left(
			\begin{array}{c}1\\\sigma \end{array}
			\right) e^{i\,\bk_j\cdot\bx}+\mbox{c.c.},\label{ExpEigAgg}
		\end{equation}
		where by c.c. we indicate the complex conjugate of the first { term in \eqref{ExpEigAgg}} and $\sigma=\dfrac{a\xi\, \sqrt{d} \, }{1 + a\xi\,\sqrt{d} \,}    	
		$.\\
		{{ The right-hand side of \eqref{EqOrd2.2} can be expressed as a linear combination of the terms \(e^0\), \(e^{i\,\mathbf{k}_j \cdot \mathbf{x}}\), \(e^{2i\,\mathbf{k}_j \cdot \mathbf{x}}\), and \(e^{i\,(\mathbf{k}_j - \mathbf{k}_l) \cdot \mathbf{x}}\). Isolating the coefficients associated with the terms \(e^{i\,\mathbf{k}_j \cdot \mathbf{x}}\), we get
				\begin{equation}
					\left(
					\begin{array}{c}R_U^j \\[2mm] R_V^j \end{array}
					\right)=\left(
					\begin{array}{c}\rho\\1 \end{array}
					\right)\frac{\pa \mathcal W_j}{\pa T_1}+\left(
					\begin{array}{c}-1\\1 \end{array}
					\right)\left[  
					\rho\,b_1 W_j  + 4\rho \left(a d  +\dfrac{
						2 \rho b_c }{a}\right)	\overline{\mathcal W}_l\,\overline{\mathcal W}_m
					\right]\,,
				\end{equation}
				with $l, m \neq j$, and $l \neq m$.\\
				It follows that the solvability condition provides
				\begin{equation}\label{Solv1}
					\left(\rho+\sigma\right)\frac{\pa \mathcal W_j}{\pa T_1}=(1-\sigma)\left[  
					\rho\,b_1 W_j + 4\rho \left(a d  +\dfrac{
						2 \rho b_c }{a}\right)		\overline{\mathcal W}_l\,\overline{\mathcal W}_m
					\right]    			, 
				\end{equation} for $j=1,2,3$.

				At this point, it is reasonable to expect the solution of \eqref{EqOrd2.2} to be of the form}
			\begin{equation}\label{Expu2}
				\begin{aligned}
					\left(
					\begin{array}{c}U_2 \\ V_2 \end{array}
					\right)
					=& \left(
					\begin{array}{c}X_0 \\ Y_0 \end{array}
					\right)\left(|\mathcal W_1|^2+|\mathcal W_2|^2+|\mathcal W_3|^2\right)\\
					&+\sum_{j=1}^3 \left(
					\begin{array}{c}\rho \\ 1 \end{array}
					\right)\,\mathcal V_j\,e^{i\,\bk_j\cdot\bx}+\sum_{j=1}^3 \left(
					\begin{array}{c}X_2 \\ Y_2\end{array}
					\right)\,\mathcal W_j^2\,e^{2\,i\,\bk_j\cdot\bx}\\
					&+\sum_{\substack{j=1,2,3 \\ l\equiv j+1\mbox{ \tiny mod }3}} \left(
					\begin{array}{c}X_1 \\ Y_1\end{array}
					\right)\,\mathcal W_j\,\overline{\mathcal W}_l\,e^{\,i\,(\bk_j-\bk_l)\cdot\bx}+\mbox{ c.c.}\,,
				\end{aligned}
			\end{equation}
			{ where here c.c. collect the complex conjugates of the explicit terms on the right-hand side.}\\
			By manipulating \eqref{EqOrd2.2}, the coefficients $X_m, Y_m$ can be obtained from the linear equations for $e^0$, $e^{i\,\bk_j\cdot\bx}$, $e^{2\,i\,\bk_j\cdot\bx}$, $e^{\,i\,(\bk_j-\bk_l)\cdot\bx}$
			\begin{equation}
				\begin{aligned}
					\left(
					\begin{array}{c}X_0 \\ Y_0 \end{array}
					\right)
					=&\left(
					\begin{array}{c}0 \\[2mm]-\dfrac{2 \rho}{a} \left({2} + \dfrac{\rho\, b_c}{a^2 d} \right) \end{array}
					\right)\\[2mm]		
					\left(
					\begin{array}{c}X_2 \\ Y_2 \end{array}
					\right)
					=& \dfrac{\rho \left( 2 a^{2} d + b_c \rho \right)}
					{4 a k_c^{2} \left( 1 - b_c + 4 k_c^{2} \right) + a^3 d \left( 1 + 4 k_c^{2} \right) \xi^{2}}
					\,	\left(
					\begin{array}{c}
						4\,k_c^2\\[2mm]
						-(1+4 \,k_c^2)\xi^2
					\end{array}
					\right)	\\[2mm]		
					\left(
					\begin{array}{c}X_1 \\ Y_1 \end{array}
					\right)
					=&\dfrac{2 \rho \left( 2 a^{2} d + b_c \rho \right)}
					{3 a k_c^{2} \left( 1 - b_c + 3 k_c^{2} \right) + a^3 d \left( 1 + 3 k_c^{2} \right) \xi^{2}}
					\,	\left(
					\begin{array}{c}
						3 k_c^2\\[2mm]		
						-(1+3 \,k_c^2)\xi^2
					\end{array}
					\right)	.
				\end{aligned}
			\end{equation}
		}
		
		{
			Lastly, we proceed to the equation at order \(\eta^3\), given in \eqref{EqOrd3}, which can be rewritten as follows
			\begin{equation}\label{EqOrd3.2}
				\mathcal L_c\left[
				\left(
				\begin{array}{c}U_3 \\ V_3 \end{array}
				\right)\right]=
				\frac{\pa}{\pa T_1}\left(
				\begin{array}{c}U_2 \\ V_2 \end{array}
				\right)+	
				\frac{\pa}{\pa T_2}\left(
				\begin{array}{c}U_1 \\ V_1 \end{array}
				\right)- \mathcal H_3\left[\left(
				\begin{array}{c}U_1 \\ V_1 \end{array}
				\right),\left(
				\begin{array}{c}U_2 \\ V_2 \end{array}
				\right)\right].
			\end{equation}
			Substituting expressions \eqref{Expu1} and \eqref{Expu2} into \eqref{EqOrd3.2}, together with the relations in \eqref{Solv1}, we can apply the Fredholm solvability condition, which yields
			\begin{equation}\label{Solv2}
				\begin{aligned}
					(\rho + \sigma)\left(\frac{\partial \mathcal V_j}{\partial T_1} + \frac{\partial \mathcal W_j}{\partial T_2}\right) =&\, 
					(1-\sigma)\left[b_2\rho W_j + b_1\rho\,\left(\mathcal V_j + 
					\dfrac{2  \rho}{a}\,\overline{\mathcal W}_m\,\overline{\mathcal W}_l\right)\right.
					\\
					&\left.+ r_1\,\left(\overline{\mathcal V}_l\,\overline{\mathcal W}_m + \overline{\mathcal V}_m\,\overline{\mathcal W}_l\right)+ r_2|\mathcal W_j|^2 + r_3(|\mathcal W_l|^2 + |\mathcal W_m|^2)\,\mathcal W_j\right],
				\end{aligned}
			\end{equation}
			for $j,l,m=1,2,3$, $j\neq l\neq m$, where the coefficients are
			\begin{equation}
				\begin{aligned}
					r_1=& 
					\dfrac{2 \rho \left( 2 a^{2} + b_c \rho \right)}{a},\\[2mm]  
					r_2=&\dfrac{2 b_c X_2 \rho}{a} + 3 \rho^{2} + 2 a \left( X_2 + (Y_0 + Y_2) \rho \right)    			
					,
					\\[2mm]
					r_3=&2 a X_1 + \dfrac{2 \left( b_c X_1 + a^{2} Y_0 + a^{2} Y_1 \right) \rho}{a} + 6 \rho^{2}    					
					.
				\end{aligned}
			\end{equation}
		}	
		
		{
			Upon combining equations \eqref{Expu1} and \eqref{Expu2}, the amplitude is recovered in its explicit expanded form}
		\begin{equation}
			{\bf A}_j=\eta\, \left(
			\begin{array}{c}\rho \\ 1 \end{array}
			\right)\,\mathcal W_j+\eta^2\, \left(
			\begin{array}{c}\rho \\ 1 \end{array}
			\right)\,\mathcal V_j+O(\eta^3),\quad j=1,2,3\,,
		\end{equation}
		where ${\bf A}_j = (A_{j}^U, A_{j}^V, A_{j}^W)^T$ denotes the amplitude vector associated with the mode ${\bf k}_j$.  
		Then we arrive at the governing equations for the amplitudes, given by
		\begin{equation}
			\frac{\pa {\bf A}_j}{\pa t}=\eta^2\, \left(
			\begin{array}{c}\rho \\ 1 \end{array}
			\right)\,\frac{\pa\mathcal W_j}{{\pa T_1}}+\eta^3\, \left(
			\begin{array}{c}\rho \\ 1 \end{array}
			\right)\,\left(\frac{\pa\mathcal W_j}{{\pa T_2}}+\frac{\pa\mathcal V_j}{{\pa T_1}}\right)+O(\eta^4),\quad j=1,2,3.
	\end{equation}}
	Accordingly, from equations \eqref{Solv1}--\eqref{Solv2}, we obtain the evolution equation governing $A_j^U$:
	\begin{equation}\label{SistAj}
		r_0\frac{\pa A_j^U}{\pa t}=\mu \,A_j^U+\left(s_1+\mu\,\tilde s_1\right)\,\overline {A_l^U}\,\overline{A_m^U}+A_j^U\,\left[s_2\,|{A_j^U}|^2+s_3\left(|{A_l^U}|^2+|A_m^U|^2\right)\right],
	\end{equation}
	being 
	\be r_0=\dfrac{\rho + \sigma}{b_c \, \rho \, (1 - \sigma)}
	,\quad \mu=\dfrac{b-b_c}{b_c},\quad s_1=\dfrac{4 a d}{b_c \rho} + \dfrac{2}{a},\quad
	\tilde s_1=\dfrac{2}{a},\quad s_i=\dfrac{r_0\,r_i}{\rho^3},\quad i=2,\,3; \label{CoefAmp}\ee
	an analogous expression holds for $A_j^V$.
	
	{We represent each amplitude in terms of its modulus and phase angle, writing \( A_j^U = \rho_j \, e^{i \phi_j} \). By substituting these expressions into system \eqref{SistAj} and equating the real and imaginary components separately, we derive the following set of equations
		\begin{equation}\label{SistRhoPhi}
			\begin{aligned}		
				r_0\, \frac{\partial \phi}{\partial t} &= -\left(s_1+\mu\,\tilde s_1\right)\frac{\rho_1^2\,\rho_2^2+\rho_2^2\,\rho_3^2+\rho_1^2\,\rho_3^2}{\rho_1\,\rho_2\,\rho_3}\,\sin(\phi) \\
				r_0\,\frac{\partial \rho_1}{\partial t} &= \mu\,\rho_1 +\left(s_1+\mu\,\tilde s_1\right) \, \rho_2 \, \rho_3 \, \cos(\phi) +s_2 \, \rho_1^3 +s_3\, (\rho_2^2 + \rho_3^2) \, \rho_1
				\\
				r_0\,\frac{\partial \rho_2}{\partial t} &= \mu \,\rho_2  +\left(s_1+\mu\,\tilde s_1\right) \, \rho_1 \, \rho_3 \, \cos(\phi) +s_2 \, \rho_2^3 +s_3\, (\rho_1^2 + \rho_3^2) \, \rho_2\\
				r_0\,\frac{\partial \rho_3}{\partial t} &=\mu \,\rho_3 +\left(s_1+\mu\,\tilde s_1\right) \, \rho_1 \, \rho_2 \, \cos(\phi) +s_2 \, \rho_3^3 +s_3\, (\rho_1^2 + \rho_2^2) \, \rho_3,
			\end{aligned}
		\end{equation}
		being $\phi=\phi_1+\phi_2+\phi_3$.
	}
	
	{The stationary solutions of system \eqref{SistRhoPhi} correspond to the distinct observable patterns. In particular, we can identify the following cases:
		\begin{itemize}
			\item[i)] Homogeneous solution with $\rho_1=\rho_2=\rho_3=0$; in this case, no pattern emerges.
			\item[ii)] Striped pattern $\mathcal S=\left(\phi,\rho_1,\,0,\,0\right)$, with $\rho_1=\sqrt{-\dfrac{\mu}{s_2}}$ ;
			\item[iii)]  Hexagonal patterns, $\mathcal H_{\bar\phi}^\pm=\left(\bar\phi,\,\bar\rho_{\pm},\,\bar\rho_{\pm},\,\bar\rho_{\pm}\right)$, with $$\bar\phi=\frac \pi2\left(1+sign\left(s_1+\mu\,\tilde s_1\right)\right),$$ $$\bar\rho_{\pm}=\frac{
				|s_1 + \mu\,\tilde s_1| \pm \sqrt{-4\,(s_2+2s_3)\,\mu + \left(s_1+\mu\,\tilde s_1\right)^2}}{2\,(s_2 + 2 s_3)};
			$$
			\item[iv)]  Mixed patterns, $\mathcal M_{\tilde\phi}=\left(\tilde\phi,\,\tilde\rho_1,\,\tilde\rho_2,\,\tilde\rho_2\right)$, with  $$\tilde\phi=\frac \pi2\left(1-sign\left(\frac{s_1+\mu\,\tilde s_1}{s_2-s_3}\right)\right),$$ $$\tilde\rho_1=\left|\frac{s_1+\mu\,\tilde s_1}{s_2-s_3}\right|,\quad
			\tilde\rho_2=\sqrt{\frac{-\mu - s_2\, \rho_1^2}{s_2+s_3}}.
			$$
		\end{itemize}	
	}
	We observe that, although the types of patterns identified for a generic value of parameter $d$ (in the range of interest) are the same as those obtained in the case $d = 1$ \cite{pena2001stability,verdasca1992reentrant}, the new parameter still plays a role by influencing the pattern amplitude.
	
\begin{figure}[ht!]
		\centering
		\includegraphics[width=0.8\linewidth]{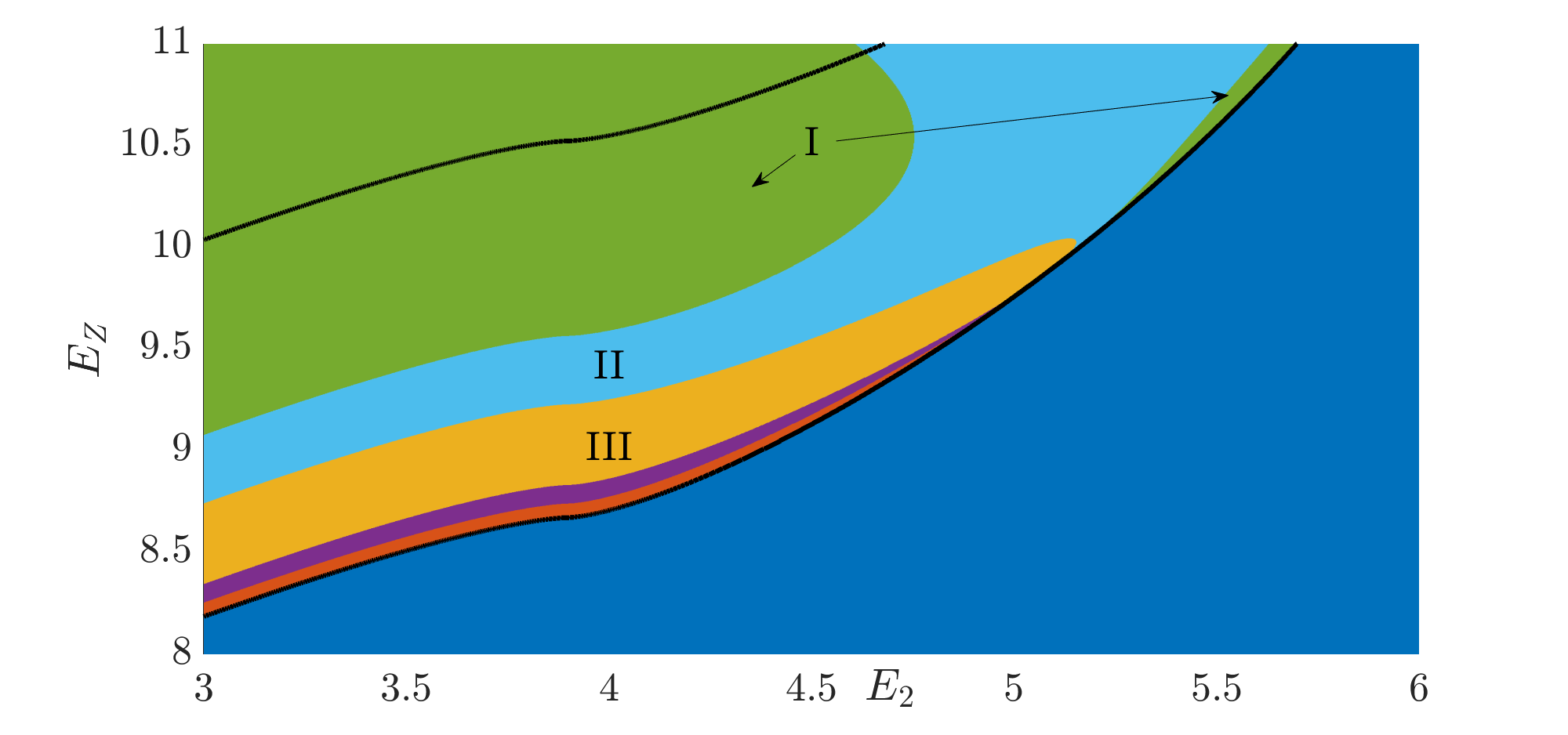}\\
		\includegraphics[width=0.8\linewidth]{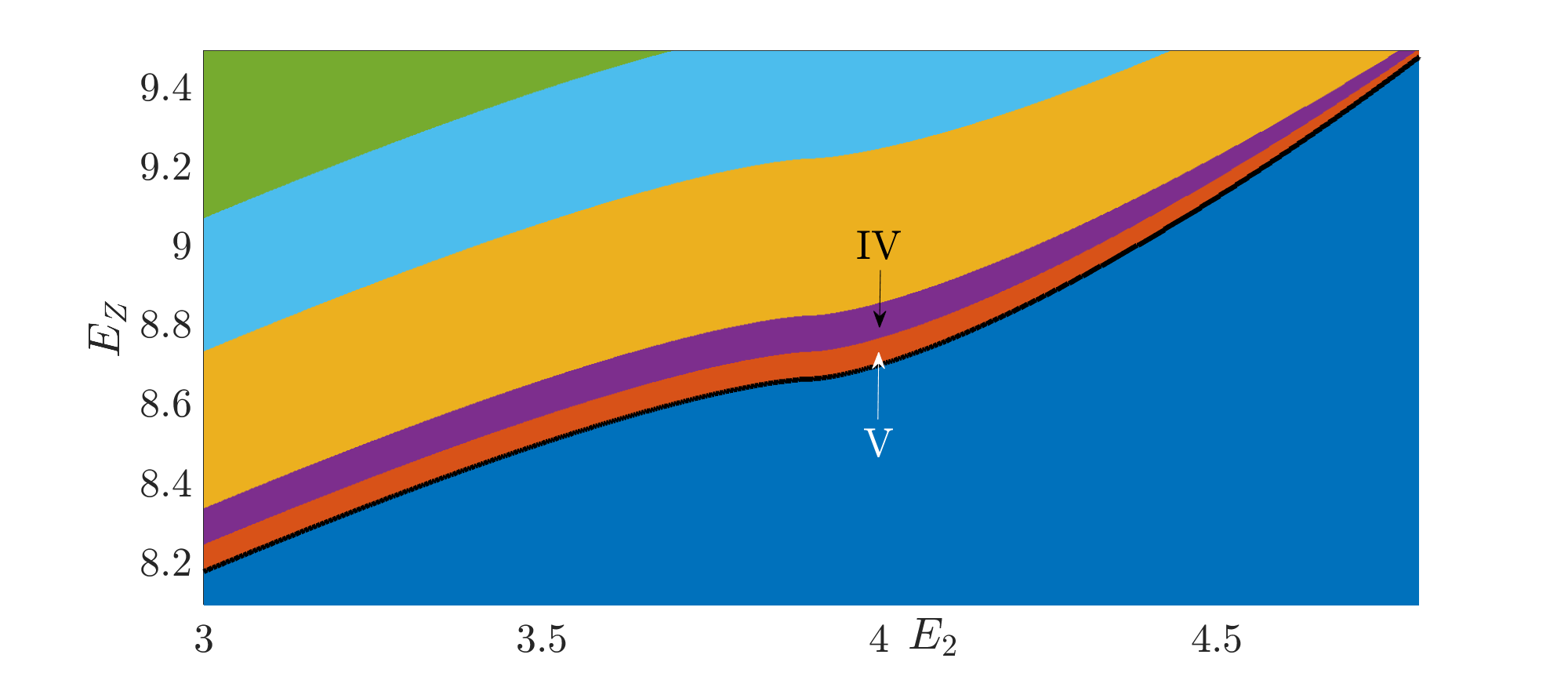}
		\caption{Values for energy levels $E_2$ and $E_Z$ relevant to the stability of the different patterns system \eqref{brusselator}, as reported in  Table \ref{tab1}. Black lines define the region where Turing instability can occur. Parameters are chosen as in \eqref{ParsBrus}.}
		\label{fig:regionPattBrus}
	\end{figure}

	To best highlight the variety of scenarios, discuss the existence and stability of the emerging patterns, and thus confirm and extend the results available in the literature \cite{pena2001stability,verdasca1992reentrant}, we choose to fix the parameters as in \eqref{ParsBrus} and let the internal energies vary. We can observe that, within the range of parameters providing Turing instabiity (region II of Figure \ref{fig:regionTurBruss}), five regions of interest can be identified, as shown in Figure \ref{fig:regionPattBrus}); in each region at least one of the patterns individuated above is stable, as reported in  Table \ref{tab1}. The thick black lines delimit region II of Figure \ref{fig:regionTurBruss}, while the blue area corresponds to region I in Figure \ref{fig:regionTurBruss}, where the equilibrium is unstable and Turing instability cannot occur.
	\begin{table}
		\centering
		\caption{Classification of steady states of system \eqref{SistRhoPhi} (stripes ${\mathcal S}$, hexagons $\mathcal{H}_l^\pm, l=0,\pi$) in each region of Figure \ref{fig:regionPattBrus}.}\label{tab1}
		\begin{tabular}{|c|c|}
			\hline
			Region & Stable Equilibria\\
			\hline
			I & $ \mathcal S,\,\mathcal{H}_0^-$\\
			\hline
			II & $\mathcal{H}_0^-$\\
			\hline
			III & $\mathcal S$\\
			\hline
			IV & $ \mathcal S,\,\mathcal{H}_\pi^-$\\
			\hline
			V & $\mathcal{H}_\pi^-$\\
			\hline     
		\end{tabular}
	\end{table}
	
	\subsection{Numerical simulations}
	{In this section we investigate the behavior of system~\eqref{brusselator} by performing numerical simulations.
		We simulate a two-dimensional square domain of length $L = 12\pi$, { discretized into a 100$\times$100 grid}.}
	{The solution is initialized by adding a random perturbation to the equilibrium values.}
	{The diffusive terms are discretized using a second-order finite-difference method, the source terms are evaluated locally, and the solution is marched in time from the initial
		conditions until convergence using a first-order forward-Euler scheme. 
		It is worth noting that once the solution is converged to a steady state, the error associated with the time discretization vanishes.
		Therefore, using a first-order time-integration scheme is acceptable for our purposes.
		{ The time step is dynamically adpated throughout the simulation to maintain numerical stability by keeping the local von Neumann number - determined from the maximum diffusion coefficient in the domain - under a given threshold.}
		At the boundary of the domain, we impose zero-gradient Neumann conditions.
		We do so by utilizing a layer of ghost cells and adapting their value as the simulation evolves. 
		The method is implemented in a modified version of the open-source Hyper2D solver \cite{boccelli2023hyper2d} tailored to reaction-diffusion systems.}
	
	At this point, we confirm the results obtained in the previous subsection by letting $E_2$ and $E_Z$ vary in the region of parameters where spatially non-homogeneous configurations are expected. We fix again parameters as in \eqref{ParsBrus} and we select three cases corresponding to different values of $E_2$ and $E_Z$; the chosen cases provide an exhaustive overview of emerging patterns. The relevant coefficients of the macroscopic system \eqref{brusselator} in each case are reported in Table \ref{tab2}.
	
	\begin{table}
		\centering
		\caption{Values of coefficients for macroscopic system \eqref{brusselator} taking microscopic values as in  \eqref{ParsBrus}, for different choices of $E_2$ and $E_Z$.}\label{tab2}
		\begin{tabular}{|c|c|c|c|c|c|c|c|c|}
			\hline
			Case & $E_2$ & $E_Z$ & $a$ & $b$ & $d$ & $D_1$ & $D_2$ & Region \\
			\hline
			1&3.81&8.68&4.53&8&2.11&1&13.33&V\\
			\hline
			2&3.78&8.76&4.53&8&1.93&1&13.33&IV\\
			\hline
			3&3.65&8.90&4.53&8&1.59&1&13.33&III\\
			\hline     
		\end{tabular}
	\end{table}
	
\begin{figure}[ht!]
		\centering
		\includegraphics[trim={0 1cm 0 1.5cm},clip,width=0.8\linewidth]{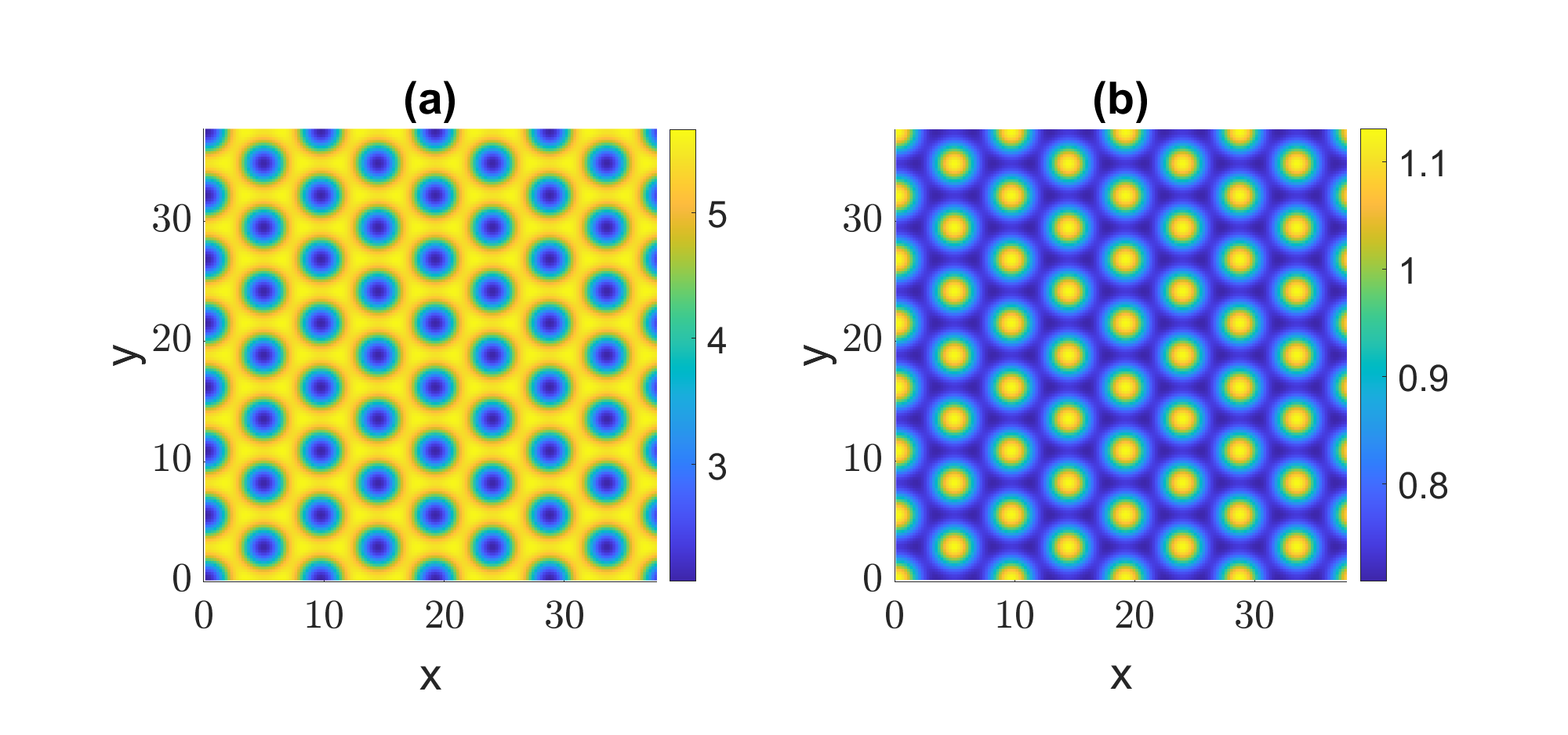}
		\caption{Pattern formation for system \eqref{brusselator}  in a squared domain $\Omega$ assuming no-flux conditions at the boundary $\partial\Omega$, and taking coefficient as reported in the first row of Table \ref{tab2}. Panel (a): density of $n_1$. Panel (b): density of $n_2$.}
		\label{fig:caso1Bruss}
	\end{figure}

	\medskip
	\textbf{Case 1.} We start by taking values of $E_2$ and $E_Z$ in region V, more precisely $E_2=3.81$ and $E_Z=8.68$ (row 1 in Table \ref{tab2}). {We run this simulation and the following ones up to final time $t=1000$.} Accordingly to the analysis, we observe the formation of hexagonal pattern corresponding to phase $\phi=\pi$, as shown in Figure \ref{fig:caso1Bruss}.

\begin{figure}[ht!]
		\centering
		\includegraphics[trim={0 1cm 0 1.5cm},clip,width=0.8\linewidth]{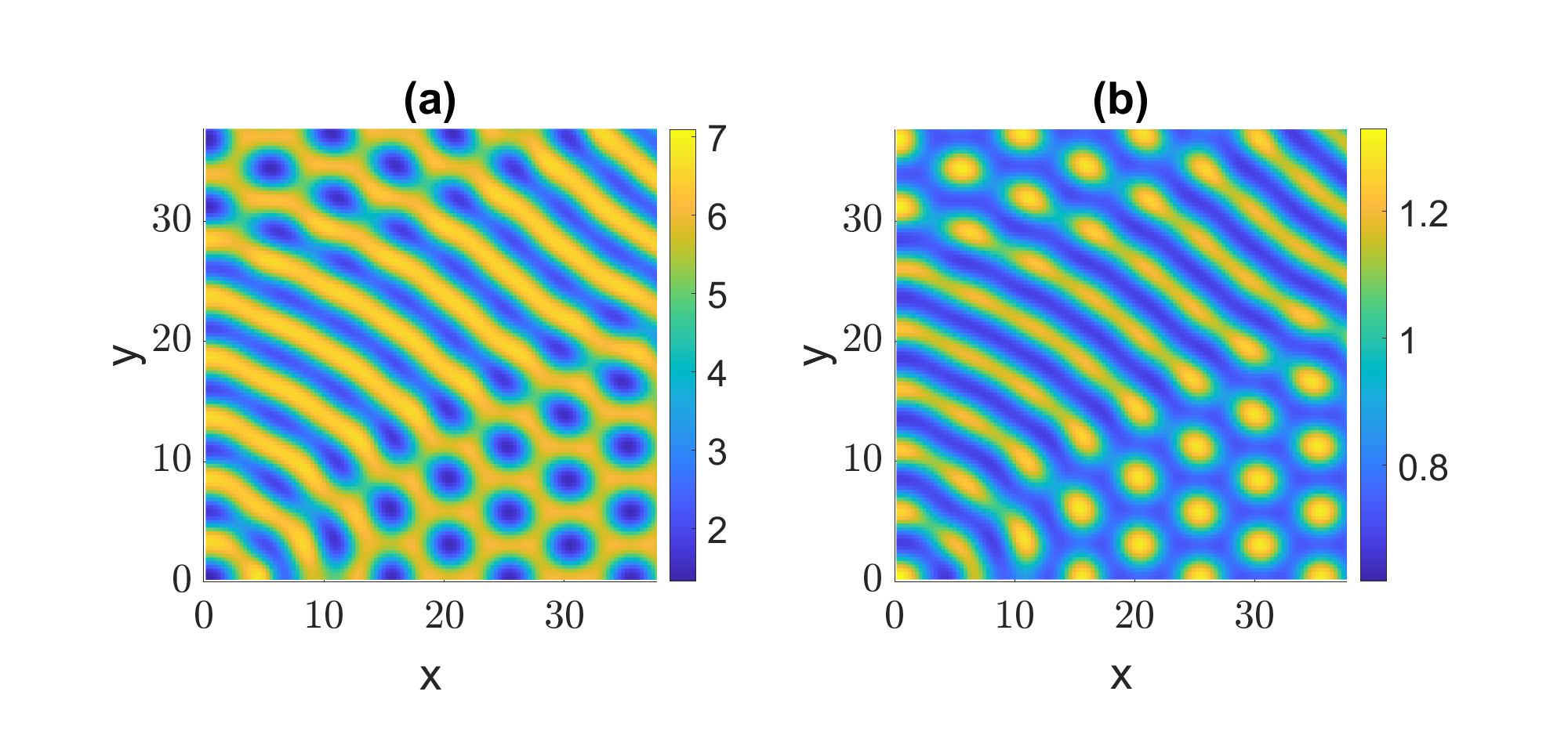}
		\caption{Pattern formation for system \eqref{brusselator}  in a squared domain $\Omega$ assuming no-flux conditions at the boundary $\partial\Omega$, and taking coefficient as reported in the second row of Table \ref{tab2}. Panel (a): density of $n_1$. Panel (b): density of $n_2$.}
		\label{fig:caso2Bruss}
	\end{figure}

	\medskip
	\textbf{Case 2.} We take values of $E_2$ and $E_Z$ in region IV, more precisely $E_2=3.78$ and $E_Z=8.76$ (row 2 in Table \ref{tab2}). In accordance with the analysis in the previous subsection, we observe the coexistence of hexagonal and striped patterns, as shown in Figure \ref{fig:caso2Bruss}.
	
\begin{figure}[ht!]
		\centering
		\includegraphics[trim={0 1cm 0 1.5cm},clip,width=0.8\linewidth]{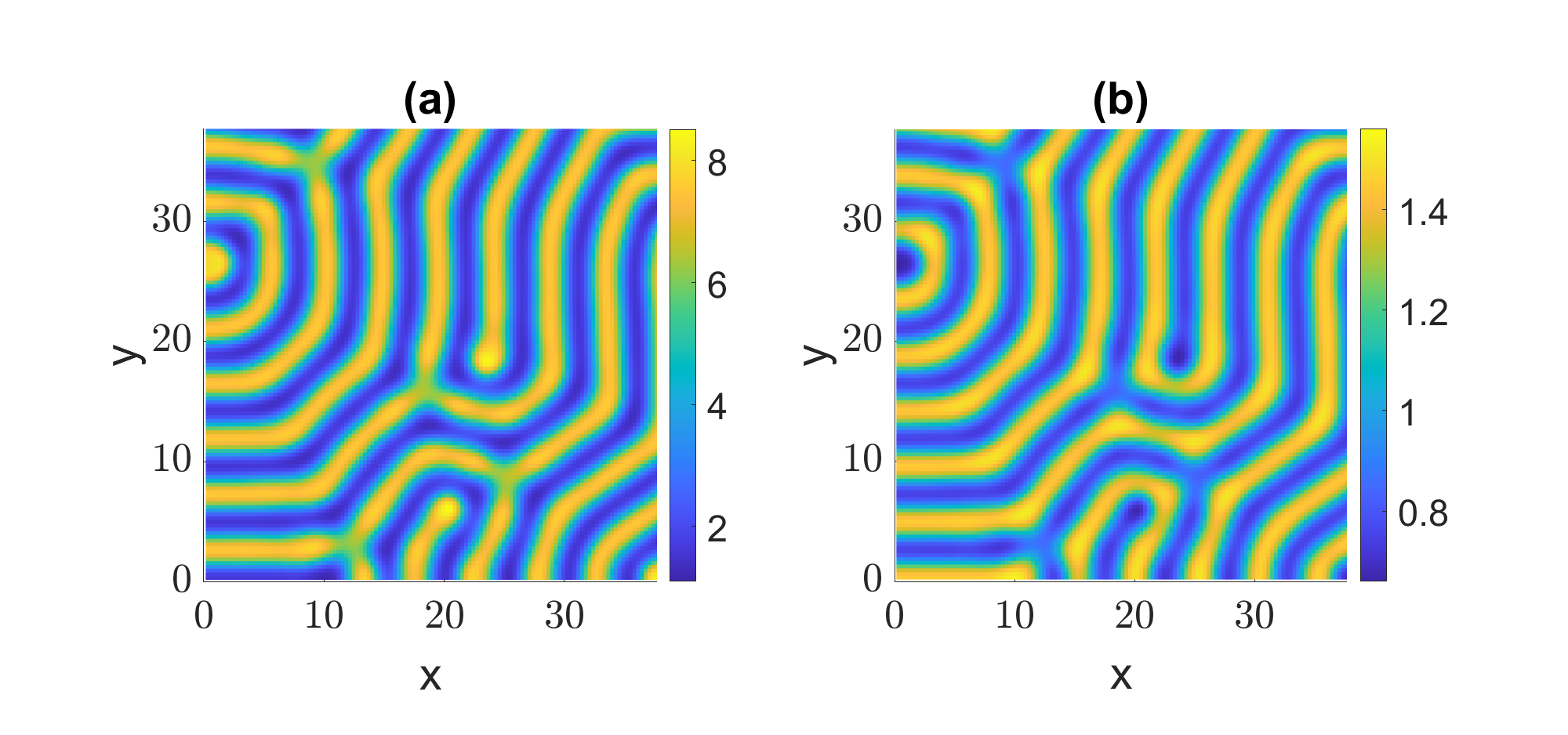}
		\caption{Pattern formation for system \eqref{brusselator}  in a squared domain $\Omega$ assuming no-flux conditions at the boundary $\partial\Omega$, and taking coefficient as reported in the third row of Table \ref{tab2}. Panel (a): density of $n_1$. Panel (b): density of $n_2$.}
		\label{fig:caso3Bruss}
	\end{figure}

	\medskip
	\textbf{Case 3.} We consider $E_2$ and $E_Z$ in region III, more precisely $E_2=3.65$ and $E_Z=8.90$ (row 3 in Table \ref{tab2}). As predicted by weakly nonlinear analysis, we observe the formation of striped pattern, as shown in Figure \ref{fig:caso3Bruss}.
	
	For the sake of brevity, we do not report the patterns in regions I and II, since their shape is similar to that of the patterns in regions IV and V, respectively; the only difference concerns the phase angle $\phi$, which passes from $\pi$ to 0, giving rise to the so-called reentrant hexagons \cite{verdasca1992reentrant}. { Although our model includes an additional parameter $d$, it reproduces qualitatively the same types of patterns reported in the literature \cite{pena2001stability}, albeit with some quantitative differences. Different parameter values, as well as different choices of the initial datum, lead to different patterns; the different ways in which instabilities interact with nonlinearities can sometimes result in the coexistence of multiple pattern types, as illustrated in Figure \ref{fig:caso2Bruss}.}
	
	%

		\section{Pattern formation in the model with nonlinear cross diffusion}
		\label{cross_analysis}
		
		We move now to system \eqref{SistCros}, with the scope to investigate the two-dimensional formation of patterns, extending the results outlined in \cite{martalo2024reaction} in the one-dimensional setting. Proceeding as in the previous case, we start by observing that the space homogeneous version of system  \eqref{SistCros} has a unique steady state
		\begin{equation}
			E=(N^*,n_Z^*)=\left( a + \frac{a c}{\zeta},\,\dfrac{\xi}{c}\right)\,,
		\end{equation}
		with  $\zeta=a^2\,d\,\beta$ and $\xi=a^2d$, whose stability can be analyzed by investigating the linearized problem for $\mathbf{U}=(N-N^*,n_Z-n_Z^*)^T$
		\begin{equation}
			\dfrac{d\,\mathbf{U}}{dt}=\mathbf{J}\mathbf{U}\,,
		\end{equation}
		where the jacobian matrix evaluated at the equilibrium is
		\begin{equation}
			\mathbf{J}=\left(
			\begin{array}{ccc}
				-\dfrac{\zeta\nu}{c+\zeta} & & c-a\dfrac{c^2\nu}{\xi(c+\zeta)}\\
				\\
				\dfrac{\zeta (-1 + \nu)}{c + \zeta} & & c\dfrac{c-\zeta}{c+\zeta}
			\end{array}\right)\,,
		\end{equation} 
		with $\nu=1+2ad$. We observe that
		\begin{equation}
			\text{tr}(\mathbf{J})=\dfrac{c^2-\zeta(c+\nu)}{c+\zeta}\text{ and }\det(\mathbf{J})=\dfrac{c\,\zeta}{c+\zeta}>0\,,
		\end{equation}
		and hence we have to require
		\begin{equation}
			\text{tr}(\mathbf{J})<0\,\Longleftrightarrow\beta>\dfrac{c^2}{\xi(c+\nu)}=:\bar{\beta}
			\label{constraint2}
		\end{equation}
		to guarantee the stability of the equilibria.
		
		\subsection{Turing instability and weakly nonlinear analysis}
\begin{figure}[ht!]
			\centering
			\includegraphics[width=0.8\linewidth]{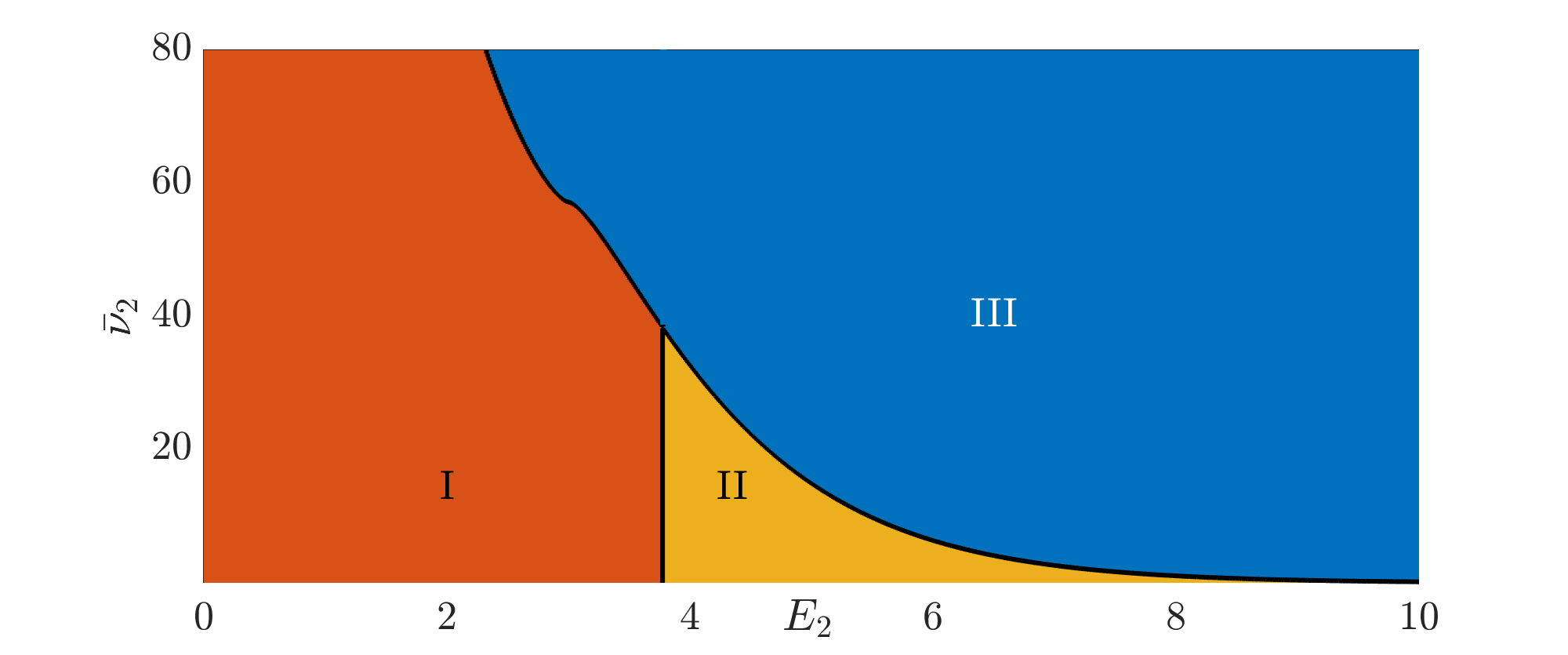}
			\caption{Values for energy level $E_2$ and collision frequency $\bar{\nu}_2$ relevant to Turing instability for system \eqref{SistCros}. In region III, condition \eqref{constraint2} is not satisfied, in region I only \eqref{constraint2} is fulfilled, while in  II both \eqref{constraint2} and \eqref{constraint2Tur} hold. Parameters are chosen as in \eqref{ParsCros}.}
			\label{fig:regionTurCross}
		\end{figure}

		In presence of diffusion, the linearized problem results
		\begin{equation}
			\dfrac{\partial\mathbf{U}}{\partial t}=\mathbf{D}\Delta_\mathbf{x}\mathbf{U}+\mathbf{J}\mathbf{U}\,,
		\end{equation}
		where the diffusion matrix is given by
		\begin{equation}
			\mathbf{D}=\left(
			\begin{array}{ccc}
				\dfrac{\zeta D_1+cD_2}{c+\zeta} & & \dfrac{c^2(D_1-D_2)}{ad(c+\zeta)}\\
				\\
				0 & & D_Z
			\end{array}\right)\,.
		\end{equation}
		As in the previous case, we look for solutions of the form given in \eqref{solutions}-\eqref{eigenfunctions}, and the Turing instability is related to the existence of at least a wavenumber $k$, such that the corresponding $\lambda_k$	has positive real part. Each $\lambda_k$ 
		is solution of the usual dispersion relation
		\begin{equation}
			\lambda^2+g(k^2)\lambda+h(k^2)=0\,,
		\end{equation}
		where
		\begin{equation}
			g(k^2)=k^2\text{Tr}\mathbf{D}-\text{Tr}\mathbf{J}>0
		\end{equation}
		because of relation \eqref{constraint2}, and
		\begin{equation}
			h(k^2)=\det(\mathbf{D})k^4+qk^2+\det(\mathbf{J})\,,
		\end{equation}
		being $q=\dfrac{c(\zeta D_1-cD_2)+\nu\zeta D_Z}{c+\zeta}$.\\
		We have to require that the minimum of $h$ attained in
		\begin{equation}
			k_c^2=-\dfrac{q}{2\det(\mathbf{D})}=-\dfrac{c(\zeta D_1-cD_2)+\nu\zeta D_Z}{2(\zeta D_1+cD_2)D_Z}
			\label{minimum_x2}
		\end{equation}
		must be negative, i.e.
		\begin{equation}
			h(k_c^2)=\dfrac{4\det(\mathbf{D})\det(\mathbf{J})-q^2}{4\det(\mathbf{D})}=\dfrac{4c\xi D_Z(\zeta D_1+cD_2)\beta-[c(\zeta D_1-cD_2)+\nu\zeta D_Z]^2}{4D_Z(c+\zeta)(\zeta D_1+cD_2)}<0\,,
		\end{equation}
		or, equivalently,
		\begin{equation}
			p(D_2)=c^4 D_2^2 
			- 2  D_2 c^2 \zeta (c D_1 + D_Z (2 + \nu)) 
			+ \zeta^2 \big((c D_1+ \nu D_Z )^2 -4 c D_1 D_Z\big) 
			>0.
		\end{equation}
		The positivity of $k_c^2$ in \eqref{minimum_x2} provides the constraint
		\begin{equation}
			D_2>\frac{\zeta \,(c D_1 + D_Z \nu)}{c^2}
			=:\hat{D_2}\,.
		\end{equation}
		We can now observe that $p(D_2)>0$ if and only if $D_2<{D_2}^-\vee D_2> D_2^+$, where  
		$$D_2^\pm=\frac{\zeta}{c^2} \Big( c D_1 + D_Z (2 + \nu) \pm 2 \sqrt{D_Z (2 c D_1 + D_Z (1 + \nu))} \Big)$$ are both positive values. Moreover, it holds 
		\begin{equation}
			p(\hat D_2)=- 4 D_Z \, \zeta^2 \, (2 c D_1 + D_Z \, \nu),
		\end{equation}
		which is negative;  
		hence, by continuity arguments, the necessary conditions for the Turing instability are fulfilled in a right neighborhood of ${D_2}^+$, and we define the critical threshold ${D_2}^c:={D_2}^+$, obtaining the constraint
		\begin{equation}
			D_2>{D_2}^c=\frac{\zeta }{c^2} \Big( c D_1 + D_Z (2 + \nu) + 2 \sqrt{D_Z (2 c D_1 + D_Z (1 + \nu))} \Big).\label{constraint2Tur}
		\end{equation}
		As in the previous case, we want to focus on the microscopic features of the model, by finding suitable sets of the microscopic parameters that lead to the conditions above to be satisfied by the macroscopic coefficients, through the relations \eqref{coefReaCros1}-\eqref{coefReaCros2}.
		{ For our purposes, we assign the following values as an illustrative example, without referring to any specific physical scenarios}
		\begin{equation}\nn
			\begin{aligned}
				&m_A = 3, \quad m_B = 5.31, \quad m_C = 5, \quad m_Y= 2.69, \quad m_Z= 0.07,
			\end{aligned}
		\end{equation}
		
		\begin{equation}\nn
			\begin{aligned}
				&\nu_{A1}^{A2} = 0.2, \quad \nu_{Z2}^{Z1} = 0.4, \quad \nu_{B1}^{AC} = 0.0001, \quad \nu_{11}^{ZB} = 0.002, \\
				&\nu_{1A} = \nu_{1B} = \nu_{1C} =  80, \\
				&\nu_{ZA} = \nu_{ZB} = \nu_{ZC} =  994.5, \\
				& n_A = 32.8,\quad n_B =2.38,\quad n_C =1,
			\end{aligned}
		\end{equation}
		
		\begin{equation}\label{ParsCros}
			E_A = 0.5, \quad E_B = 0.5, \quad E_C = 0.5, \quad E_1 = 3, \quad E_Z = 1.5;
		\end{equation}
		consequently, we reduce to discuss the parameter range of interest in terms of the energy level $E_2$ and of the collision frequencies of the polyatomic component with the background $\nu_{2A} = \nu_{2B} = \nu_{2C}=:\bar\nu_2$, in such a way that the macroscopic coefficients satisfy conditions \eqref{constraint2}-\eqref{constraint2Tur}. In Figure \ref{fig:regionTurCross}, we indivduate three regions in the $E_2-\bar\nu_2$ plane: in region III, condition \eqref{constraint2} does not hold, thus the spatially homogeneous equilibrium is unstable and patterns cannot form; in region I only \eqref{constraint2} is fulfilled, while in  II both \eqref{constraint2} and \eqref{constraint2Tur} are satisfied. Therefore, the emergence of spatial patterns must be investigated in region II. 
		
		As done before, a weakly nonlinear analysis of the problem is performed; then, a third order Taylor expansion around the equilibrium system $\left(N^*,n^*_Z\right)$ leads to
		
		\begin{equation}\label{SistComp1Cr}
			\begin{aligned}
				\frac{\partial {\bf U}}{\partial t} &= \mathcal L\,{\bf U} + \mathcal H[{\bf U}], \quad \mbox{for} \quad
				{\bf U}=\left(\begin{array}{c}U\\V\end{array}\right)=\left(\begin{array}{c}N- N^*\\n_Z-n_Z^*\end{array}\right),
			\end{aligned}
		\end{equation}
		with $\mathcal L  ={{\mathbf D}}\Delta_{\bf x}+ {{\mathbf J}}$, being 
		$ {{\mathbf J}}$ and  ${{\mathbf D}}$  the Jacobian  and the diffusion matrix, respectively, while the nonlinear part of the expansion $
		\mathcal H[{\bf U}] = \left(\mathcal H^U[{\bf U}],\,\mathcal H^V[{\bf U}] \right)^T$
		has the following components
		\begin{equation}
			\begin{aligned}
				\mathcal H^U[{\bf U}]=
				&-\frac{3 \beta^2 c^4 \delta \, \Delta V \, V^2}{a d (c+\zeta)^3}
				+ \frac{\beta c^4 V^3 (-a \beta + 2 c - 2 \zeta)}{\xi (c+\zeta)^3}
				+ \frac{2 \beta c^3 \delta \, \Delta V \, V}{a d (c+\zeta)^2}
				+ \frac{2 \beta c^3 \delta (\nabla V)^2}{a d (c+\zeta)^2} \\
				&+ \frac{\beta c^3 U V^2 (a \beta - 2 c + 4 \zeta)}{a (c+\zeta)^3}
				+ \frac{c^3 V^2 (a \beta - c + 2 \zeta)}{\xi (c+\zeta)^2}
				- \frac{c^2 \delta \, \Delta V}{a d (c+\zeta)}
				\\
				&- \frac{\zeta U (a + 2 \xi)}{a (c+\zeta)}
				- \frac{c V \left(a \beta c + \zeta (c - \zeta)\right)}{\zeta (c+\zeta)}
				+ \frac{2 \beta c^3 \delta \, \Delta V \, \zeta U V}{\xi (c+\zeta)^3}
				+ \frac{4 \beta c^3 \delta \, \zeta V \nabla U \nabla V}{\xi (c+\zeta)^3} \\
				&+ \frac{\beta^2 c^3 \delta \, \Delta U \, V^2}{(c+\zeta)^3}
				- \frac{2 \beta c^2 d \zeta U^2 V}{(c+\zeta)^3}
				- \frac{c^2 \delta \, \Delta V \, \zeta U}{\xi (c+\zeta)^2}
				- \frac{2 c^2 \delta \, \zeta \nabla U \nabla V}{\xi (c+\zeta)^2} \\
				&- \frac{\beta c^2 \delta \, \Delta U \, V}{(c+\zeta)^2}
				- \frac{d \zeta^2 U^2}{(c+\zeta)^2},
			\end{aligned}
		\end{equation}
		and
		
		\begin{equation}
			\begin{aligned}
				\mathcal H^V[{\bf U}]
				=&\frac{d U^2 \zeta^2}{(c + \zeta)^2} 
				+ \frac{2 a c^3 d U V^2 \beta (c - 2 \zeta)}{(c + \zeta)^3 \, \xi} 
				+ \frac{2 c^2 d U^2 V \zeta^2}{(c + \zeta)^3 \, \xi} \\
				&+ \frac{2 c^4 V^3 \beta (-c + \zeta)}{(c + \zeta)^3 \, \xi} 
				+ \frac{c^3 V^2 (c - 2 \zeta)}{(c + \zeta)^2 \, \xi} 
				+ \frac{4 a c^2 d U V \zeta}{(c + \zeta)^2 \, \xi},
			\end{aligned}
		\end{equation}
		with $\delta=D_2-D_1$.
		
		At this point, the diffusion coefficient $D_2$ can be expanded around the critical threshold $D_2^c$ in terms of a small parameter $\eta$
		\begin{equation}
			D_2=  D_2^c +\eta\,D_2^1+\eta^2\,D_2^2+\eta^3\,D_2^3+O(\eta^3);
		\end{equation}
		analogously, we expand also the solution vector ${\bf U}$  in terms of $\eta$ as in \eqref{UExp} and rely on the multiple time scale by means of \eqref{timeExp}. When substituting the expansions \eqref{UExp}–\eqref{timeExp} into system \eqref{SistComp1Cr} and collecting terms of the same order in $\eta$, we get a set of equations analogous to \eqref{EqOrd1}-\eqref{EqOrd3}. \\
		In detail, at order $\eta$, we obtain the equation 
		\begin{equation}\label{SistOrd1Cr}
			\mathcal L_c\left[
			\left(
			\begin{array}{c}U_1 \\ V_1 \end{array}
			\right)\right]=	\left(
			\begin{array}{c}\mathcal L_c^U[U_1,V_1] \\[2mm]\mathcal L_c^V[U_1,V_1] \end{array}
			\right)
			=0,
		\end{equation}
		where $\mathcal L_c$
		has entries
		\begin{equation}
			\begin{aligned}
				\mathcal L_c^U[U_1,V_1] &=-\frac{c^2 \delta^c \, \Delta V_1}{a d (c+\zeta)}
				-\frac{c V_1 \left(c + a c d - a d \zeta \right)}{a d (c+\zeta)}
				+ \Delta U_1 \left(D_2^c - \frac{\delta^c \, \zeta}{c+\zeta}\right)
				- \frac{U_1 \zeta (a + 2 \xi)}{a (c+\zeta)},
			\end{aligned}
		\end{equation}
		
		\begin{equation}
			\begin{aligned}
				\mathcal L_c^V[U_1,V_1] &=
				D_Z \, \Delta V_1 
				+ \frac{c V_1 (c - \zeta)}{c+\zeta} 
				+ \frac{2 a d U_1 \zeta}{c+\zeta},
			\end{aligned}
		\end{equation}
		with $\delta_c=D_2^c-D_1$. By solving the equation, we can write the solution in the form \eqref{Expu1} with
		$$
		\rho = 
		\frac{
			c \left( c \left( -a d + \delta^c k_c^2 - 1 \right) + a d \zeta \right)
		}{	a d \left( k_c^2 (c D_2^c + \zeta (D_2^c - \delta^c)) + \zeta \right) + 2 d \zeta \xi.
		}
		$$
		
		At order $\eta^2$, we obtain again 
		\begin{equation}\label{EqOrd2.2Cr}
			\mathcal L_c\left[
			\left(
			\begin{array}{c}U_2 \\ V_2 \end{array}
			\right)\right]
			= 
			\frac{\pa}{\pa T_1}\left(
			\begin{array}{c}U_1 \\ V_1 \end{array}
			\right)
			- \mathcal H_2\left[\left(
			\begin{array}{c}U_1 \\ V_1 \end{array}
			\right)\right]
			,\quad
			\mathcal H_2\left[\left(
			\begin{array}{c}U_1 \\ V_1 \end{array}
			\right)\right]=\left(
			\begin{array}{c}\mathcal H_2^U\left[U_1,V_1\right] \\[2mm] \mathcal H_2^V\left[U_1,V_1\right]  \end{array}
			\right),
		\end{equation} and the components of operator $\mathcal H_2
		\left[
		U_1, V_1
		\right]$ are given in \eqref{H_2_U} and \eqref{H_2_V}. Also in this case, we  apply the solvability condition, being the eigenvectors of $\mathcal L_c^+$ of the form \eqref{ExpEigAgg}, with
		$$\sigma = 
		\frac{
			c \left( c \left( -a d + \delta^c k_c^2 - 1 \right) + a d \zeta \right)
		}{	a d \left( -c^2 + c \left( D_Z k_c^2 + \zeta \right) + D_Z \zeta k_c^2 \right)
		}
		.$$ The condition provides and expression analogous to \eqref{Solv1}, that we omit for brevity. Afterward, we can write again the solution of \eqref{EqOrd2.2Cr} in the form \eqref{Expu2}, where the coefficients are given explicitly in the appendix \ref{appendix}.
		
		At order $\eta^3$, in the equation 
		\begin{equation}\label{EqOrd3.2Cr}
			\begin{aligned}
				\mathcal L_c\left[
				\left(
				\begin{array}{c}U_3 \\ V_3 \end{array}
				\right)\right]=&
				\frac{\pa}{\pa T_1}\left(
				\begin{array}{c}U_2 \\ V_2 \end{array}
				\right)+	
				\frac{\pa}{\pa T_2}\left(
				\begin{array}{c}U_1 \\ V_1 \end{array}
				\right)- \mathcal H_3\left[\left(
				\begin{array}{c}U_1 \\ V_1 \end{array}
				\right),\left(
				\begin{array}{c}U_2 \\ V_2 \end{array}
				\right)\right],
				\\[2mm]
				&\mathcal H_3\left[\left(
				\begin{array}{c}U_1 \\ V_1 \end{array}
				\right),\left(
				\begin{array}{c}U_2 \\ V_2 \end{array}
				\right)\right]= \left(
				\begin{array}{c}\mathcal H_3^U\left[
					U_1, V_1, U_2,V_2\right]\\[2mm]
					\mathcal H_3^V\left[
					U_1, V_1, U_2,V_2\right]
				\end{array}	
				\right),
			\end{aligned}
		\end{equation} the operator  $\mathcal H_3
		\left[
		U_1, V_1, U_2,V_2
		\right]$, whose components are given in \eqref{H_3_U} and \eqref{H_3_V}, intervenes; applying again the solvability condition to such equation, we obtain an expression analogous to \eqref{Solv2}, that we omit for brevity.
		
		Finally, following the technique outlined in the previous section, we are able to recover the equations for the amplitudes
		\begin{equation}\label{SistAj_2}
			r_0\frac{\pa A_j^U}{\pa t}=\mu \,A_j^U+\left(s_1+\mu\,\tilde s_1\right)\,\overline {A_l^U}\,\overline{A_m^U}+A_j^U\,\left[s_2\,|{A_j^U}|^2+s_3\left(|{A_l^U}|^2+|A_m^U|^2\right)\right],
		\end{equation}
		where
		$$
		r_0=\frac{a d (c + \zeta)(\rho + \sigma)}{c D_2^c k_c^2 \,(c - a d \rho)},\quad \mu=\dfrac{D_2-D_2^c}{D_2^c}, 
		$$
		and $s_1$, $\tilde s_1$, $s_2$, $s_3$ are given in \eqref{CoefSistAj_2}.
		
		\begin{figure}[ht!]
			\centering
			\includegraphics[width=0.8\linewidth]{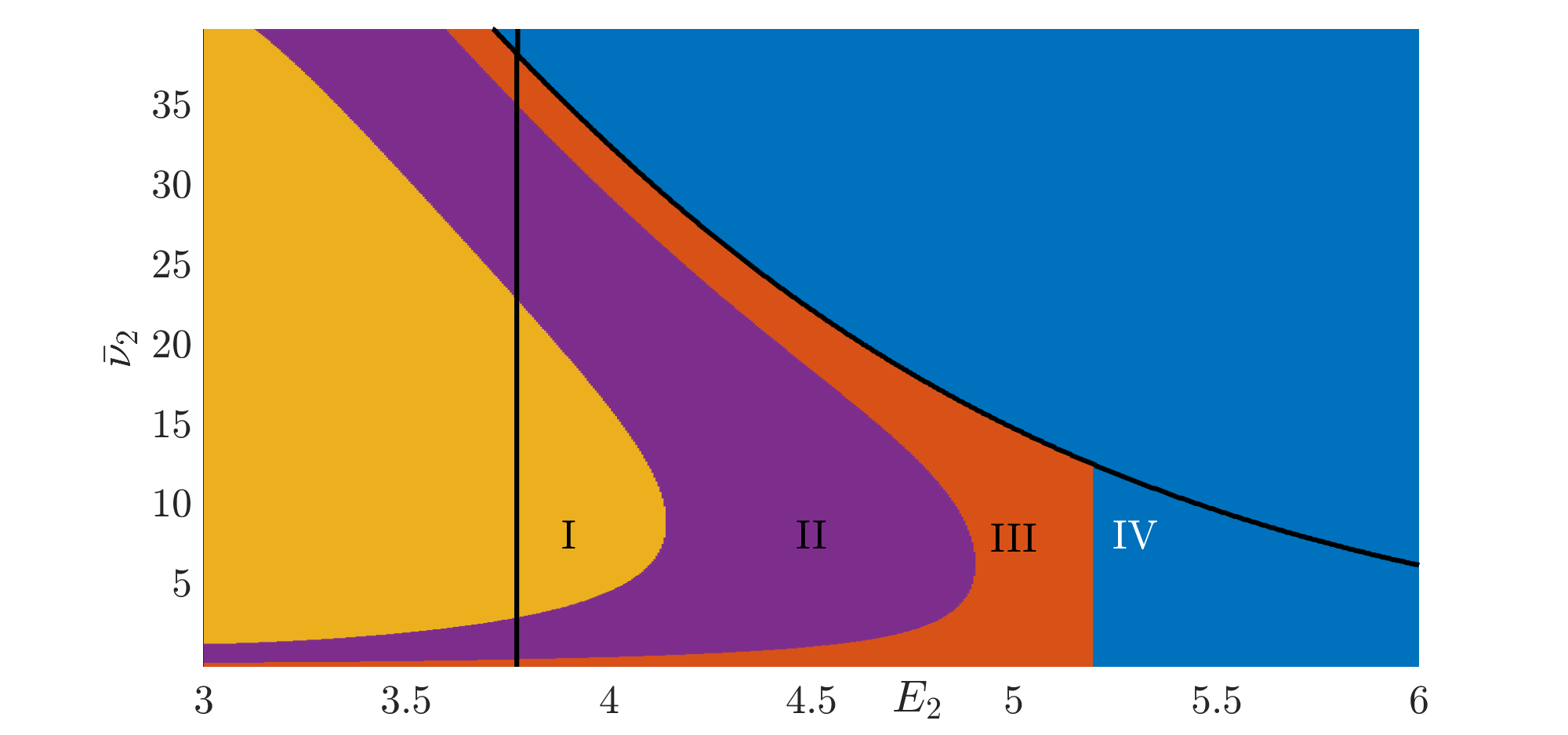}
			\caption{Values for energy level $E_2$ and collision frequency $\bar\nu_2$ relevant to the stability of the different patterns for system \eqref{SistCros}. Black lines define the region where Turing instability can occur, leading to stable striped patterns (region I),  coexistence of stable striped and hexagonal pattern (region II), stable hexagonal pattern  (region III) or to stable pattern-less configuration (region IV). Parameters are chosen as in \eqref{ParsBrus}.}
			\label{fig:regionWeakCross}
		\end{figure}
		As above, we can discuss the emergence and stability of patterns in terms of microscopic quantities; more precisely, in this case, we fix some parameters as in \eqref{ParsCros} and let $E_2$ and $\bar\nu_B$ vary. In Figure \ref{fig:regionWeakCross}, different regions are individuated, according to the stability of the pattern: we obtain striped patterns in region I, hexagons in region III, an the coexistence of both in region II.  We highlight that no patterns are expected to exist and to be stable in region IV.
		Results are summarized in Table \ref{tab_new}.
		\begin{table}
			\centering
			\caption{Classification of steady states in cross diffusion model (stripes ${\mathcal S}$, hexagons $\mathcal{H}_l^\pm, l=0,\pi$) in each region of Figure \ref{fig:regionWeakCross}.}\label{tab_new}
			\begin{tabular}{|c|c|}
				\hline
				Region & Stable Equilibria\\
				\hline
				I & $ \mathcal S$\\
				\hline
				II & $ \mathcal S,\,\mathcal{H}_\pi^-$\\
				\hline
				III & $\mathcal{H}_\pi^-$\\
				\hline
				IV & --\\
				\hline    
			\end{tabular}
		\end{table}

		\subsection{Numerical simulations}
		The theoretical results obtained in the previous subsection can be confirmed by means of numerical simulations for system \eqref{SistCros} by taking $E_2$ $\bar\nu_2$ in the different regions of Figure \ref{fig:regionWeakCross}; the corresponding coefficients of the macroscopic system \eqref{SistCros} are reported in Table \ref{tab3}. All the remaining microscopic parameters are set as in \eqref{ParsCros}. { In this case,  numerical simulations are performed on a two-dimensional square domain of length $L = 10\pi$,  discretized into a 250$\times$250 grid, starting from a perturbation of the homogeneous equilibrium shaped as cosines in the three directions $\mathbf{k}_i$, up to final time $t=500$.}
		\begin{table}
			\centering
			\caption{Values of coefficients for macroscopic system \eqref{SistCros} taking microscopic values as in  \eqref{ParsCros}, for different choices of $E_2$ and $\bar\nu_2$.}\label{tab3}
			\begin{tabular}{|c|c|c|c|c|c|c|c|c|c|c|}
				\hline
				Case & $E_2$ & $\bar\nu_2$ & $a$ & $\beta$ & $c$ & $d$ & $D_1$ & $D_2$ & $D_Z$ & Region\\
				\hline
				1&4.04&13.34&2.5&0.11&40&20&1&6&2 & I\\
				\hline
				2&4.45&13.34&2.5&0.15&40&20&1&6&2 & II\\
				\hline
				3&4.98&13.34&2.5&0.23&40&20&1&6&2 & III\\
				\hline     
			\end{tabular}
		\end{table}
		
\begin{figure}[ht!]
			\centering
			\includegraphics[width=0.8\linewidth]{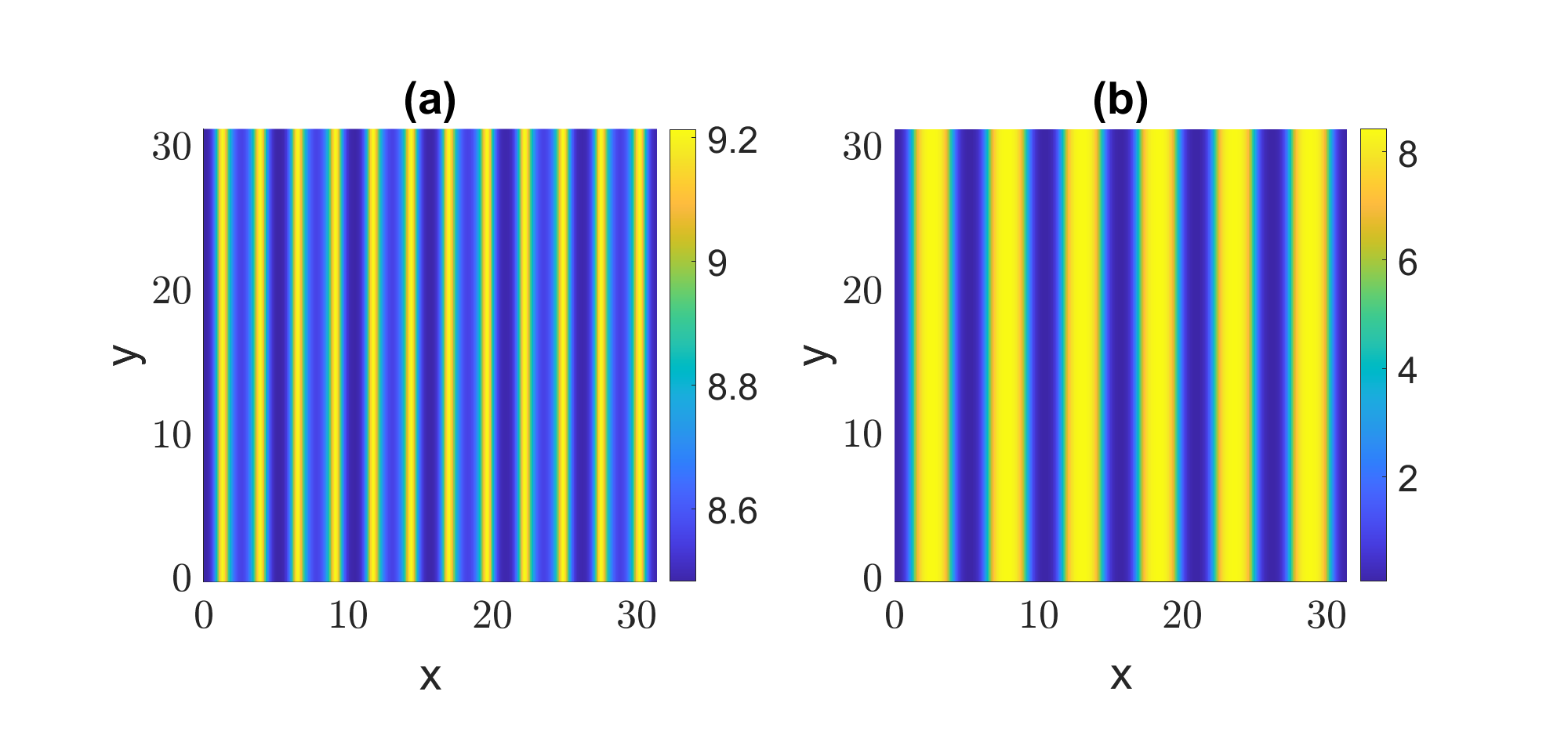}
			\includegraphics[width=0.8\linewidth]{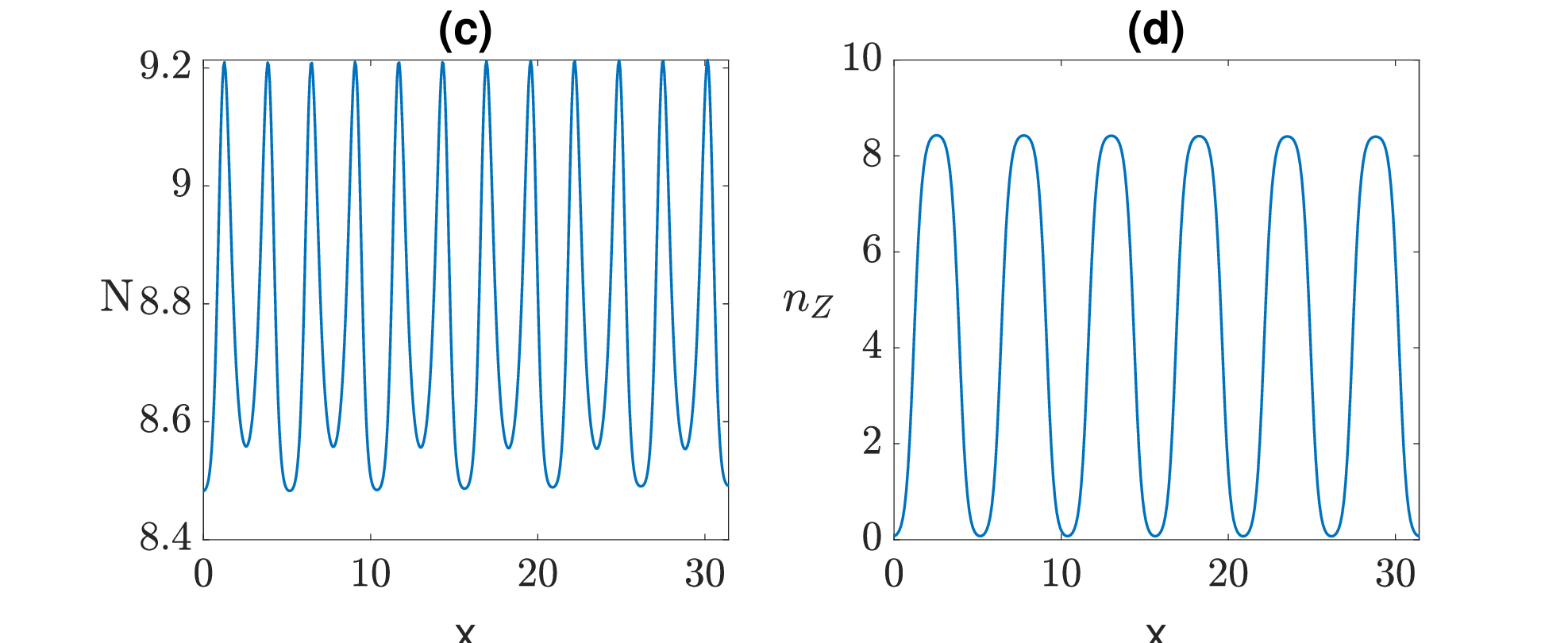}
			\caption{Pattern formation for system \eqref{SistCros}  in a squared domain $\Omega$ assuming no-flux conditions at the boundary $\partial\Omega$, and taking coefficient as reported in the first row of Table \ref{tab3}. Panel (a): density of $N$. Panel (b): density of $n_Z$. Panel (c): section of density of $N$ at $y=15$. Panel (d): section of density of $n_Z$ at $y=15$. }
			\label{fig:caso1Cross}
		\end{figure}

		\medskip
		\textbf{Case 1.} As first case, we take values of $E_2$ and $\bar\nu_2$ in region I (more precisely $E_2=4.04$ and $\bar\nu_2=13.24$). As predicted by the analysis, this choice leads to stable striped pattern, as shown in Figure \ref{fig:caso1Cross}.

\begin{figure}[ht!]
			\centering
			\includegraphics[width=0.8\linewidth]{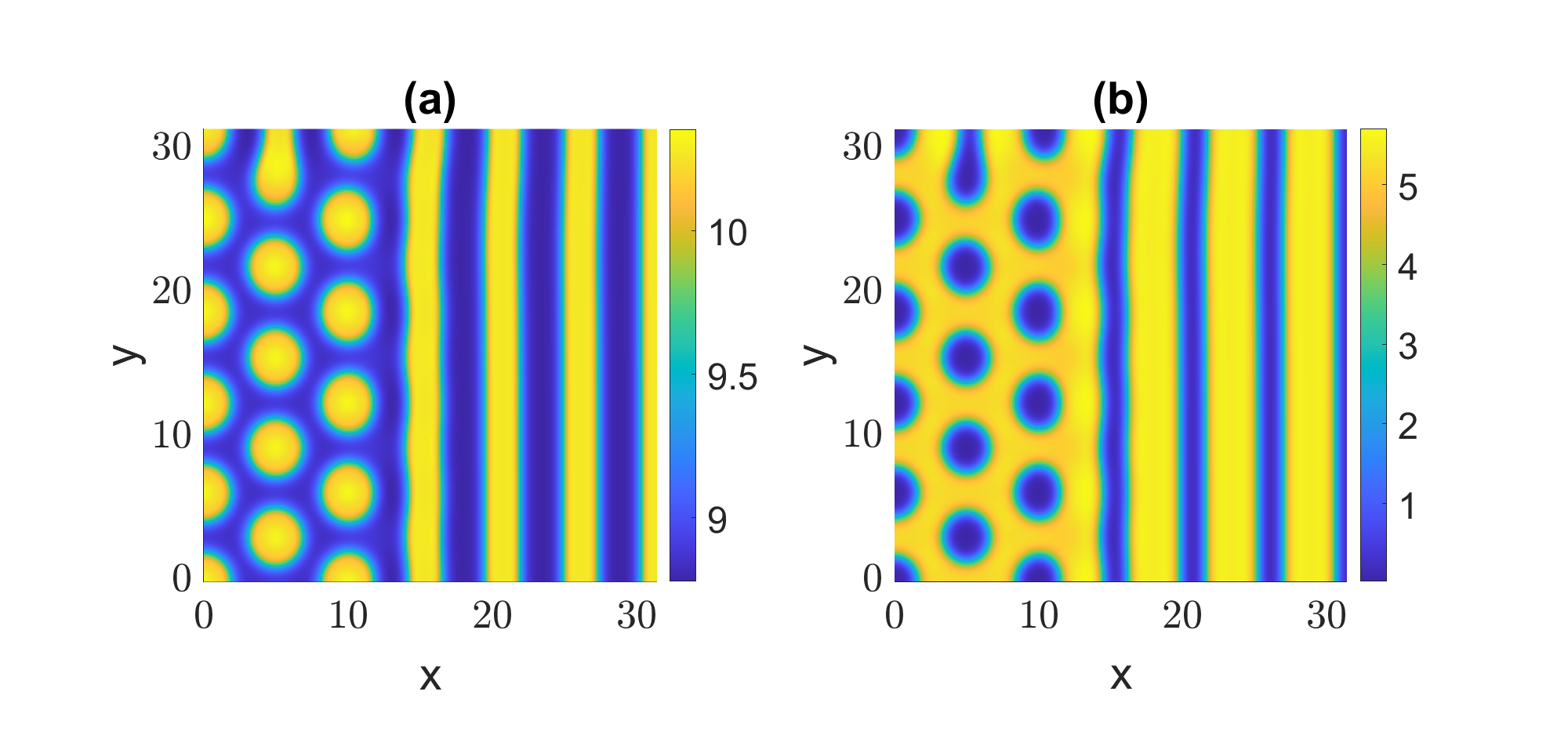}
			\includegraphics[width=0.8\linewidth]{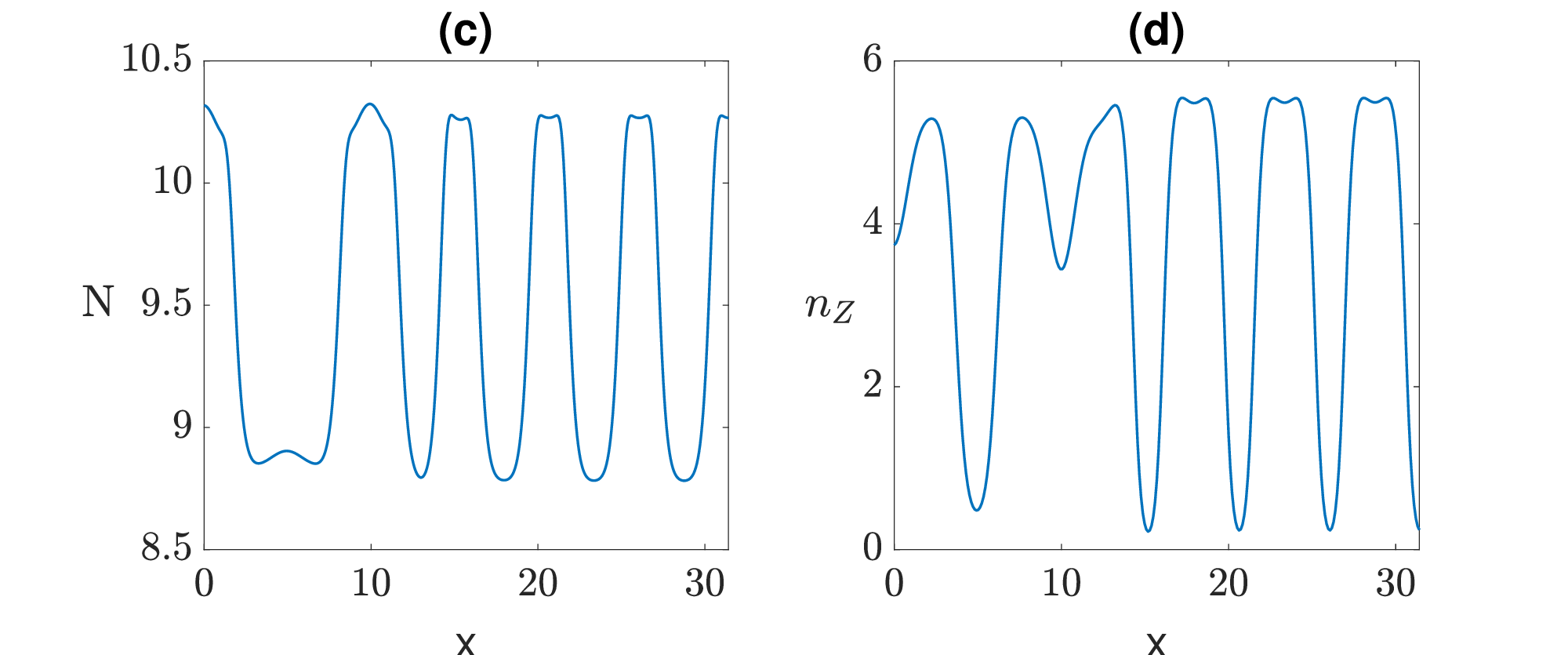}
			\caption{Pattern formation for system \eqref{SistCros}  in a squared domain $\Omega$ assuming no-flux conditions at the boundary $\partial\Omega$, and taking coefficient as reported in the second row of Table \ref{tab3}. Panel (a): density of $N$. Panel (b): density of $n_Z$. Panel (c): section of density of $N$ at $y=15$. Panel (d): section of density of $n_Z$ at $y=15$. }
			\label{fig:caso2Cross}
		\end{figure}

		\medskip
		\textbf{Case 2.} As second test, we consider $E_2$ and $\bar\nu_2$ in region II ($E_2=4.04$ and $\bar\nu_2=23.19$). As shown, these values correspond to the coexistence of stable hexagons (with phase angle $\phi=\pi$) and striped patterns, as shown in Figure \ref{fig:caso2Cross}.
		
\begin{figure}[ht!]
			\centering
			\includegraphics[width=0.8\linewidth]{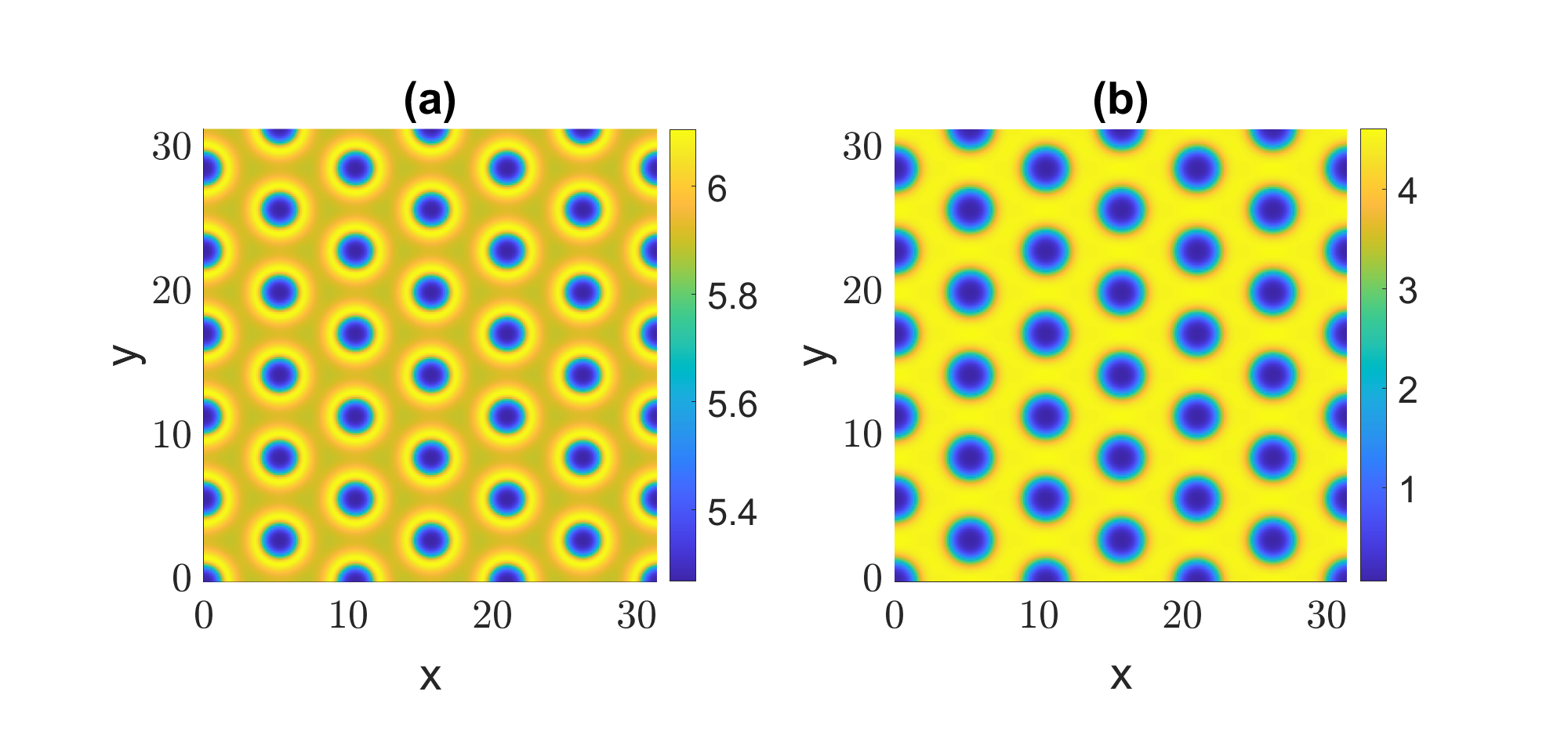}
			\includegraphics[width=0.8\linewidth]{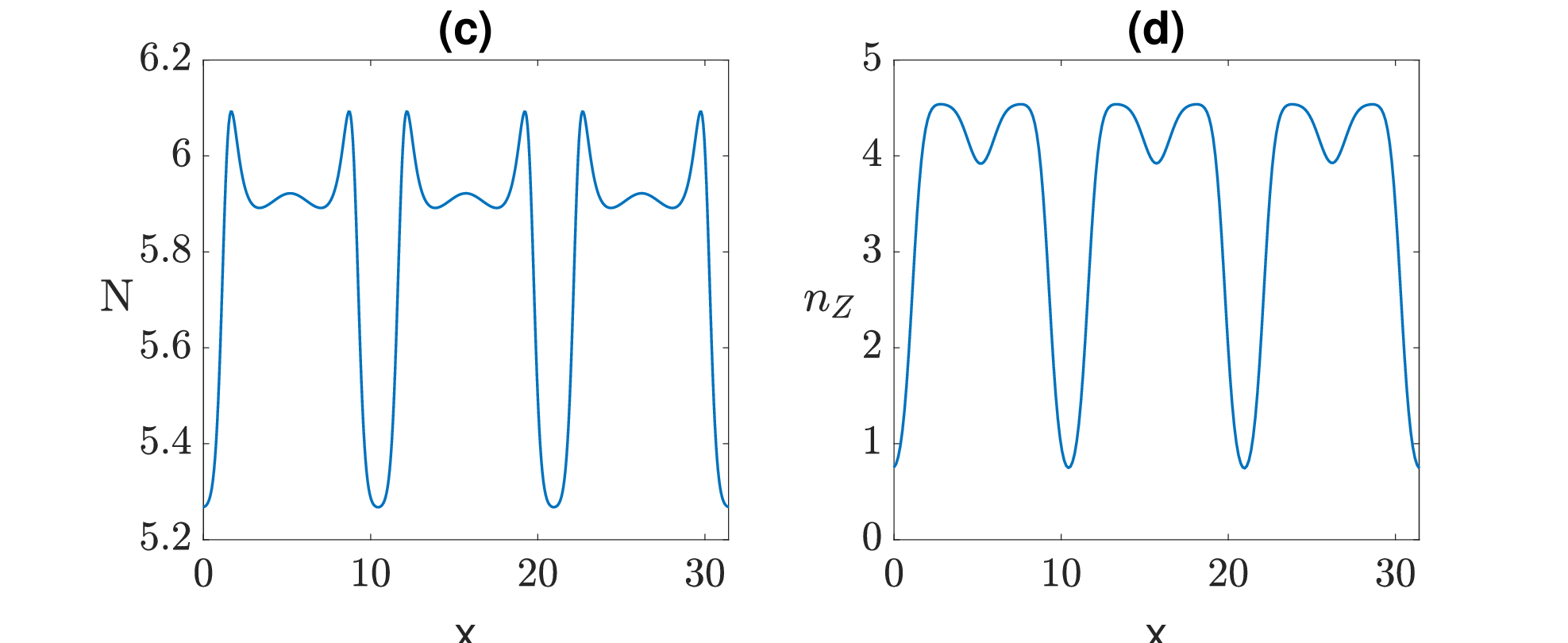}
			\caption{Pattern formation for system \eqref{SistCros}  in a squared domain $\Omega$ assuming no-flux conditions at the boundary $\partial\Omega$, and taking coefficient as reported in the third row of Table \ref{tab3}. Panel (a): density of $N$. Panel (b): density of $n_Z$. Panel (c): section of density of $N$ at $y=15$. Panel (d): section of density of $n_Z$ at $y=15$. }
			\label{fig:caso3Cross}
		\end{figure}

		\medskip
		\textbf{Case 3.} Finally, we choose $E_2=4.98$ and $\bar\nu_2=13.34$, i.e. in region III. The emergence of stable hexagons corresponding (with phase angle $\phi=\pi$) is confirmed, as shown in Figure \ref{fig:caso3Cross}.

		{ Although some patterns presented here may resemble those reported in other contexts, it is worth underlying that the results shown are entirely new; as already noted, the proposed model reproduces the diffusive mechanism of \cite{conforto2018reaction}, but its reactive terms are substantially different.\\
		Moreover,} we point out that, as shown also in the one-dimensional case in \cite{martalo2024reaction}, the numerical results suggest that the profile of $N$ is sum of more than one mode in each direction $\mathbf{k}_i$. Therefore, a deeper analysis of the solution near the critical threshold is needed; this will be subject of future investigation. 
		
		\section{Conclusions}
		\label{conclusions}
		In this paper, we have discussed Turing instability and pattern formation in two-dimensional domains for two reaction-diffusion models for gas mixtures of monatomic and polyatomic components, obtained as diffusive limits of kinetic equations. The derivation from the mesoscopic level offers the significant advantage of relating macroscopic coefficients, usually set empirically, to microscopic parameters, and therefore to the mechanisms of interaction among the components or with the environment. This connection also rigorously identifies suitable ranges for selecting the macroscopic parameters.
		
		In the first model under consideration, we showed how the kinetic derivation of Brusselator-type equations reveals the presence of an additional parameter $d$, which is typically set equal to 1 in the literature. The presence of this parameter has several effects, as it influences, for instance, the equilibria and their stability. With respect to pattern formation, $d$ does not affect the variety of patterns that may emerge (which remain the same as in the case $d = 1$), but it does affect their amplitude.
		
		In the second framework analyzed, the resulting model exhibits nonlinear cross-diffusion terms, as in \cite{conforto2018reaction}, though with different reactive terms. Pattern formation is investigated theoretically through weakly nonlinear analysis and confirmed numerically, thereby extending the preliminary results obtained in the one-dimensional case in \cite{martalo2024reaction}.
		
		The proposed analysis shows the strength of these tools in linking macroscopic dynamics to microscopic interaction mechanisms. These, along with other approaches (such as active particle theory or redistribution operators), allow one to derive, and thus mathematically justify, several models already known in the literature, while providing broader insight into their behavior. { In this perspective, future work will be devoted to the derivation of alternative reactive terms, such as those proposed in \cite{conforto2018reaction}. A careful calibration of these terms, based also on a deep knowledge of the underlying microscopic interaction mechanisms, is essential, as the intrinsic nonlinearities in them play a crucial role in enhancing the realism of the models. These nonlinear effects not only allow a more faithful representation of the processes but also provide the key element to reproducing and interpreting the rich variety of spatial patterns observed in real systems.}
		
		Finally, as argued above, further attention will be devoted to analyze the behavior of the densities and the possible presence of interacting modes{  whose study is particularly useful in applications because the coupling among spatial modes can lead to the emergence of complex patterns, influene their stability, and affect their shape. Understanding these interactions could help to better clarify the mechanisms behind complex structures observed in real ecosystems, such as mixed and oscillating patterns, and to interpret empirical data and evidence. In addition, in this work, we have focused on the patterns predicted analytically near the critical thresholds, showing that stable spatial structures emerge from perturbations of the equilibrium. More complex and temporally oscillating patterns may arise from initial conditions far from these thresholds \cite{bisi2025derivation}, and we plan to investigate these richer dynamics in future work.}
		
		\vspace{1cm}
		
		\emph {Acknowledgments} 
		This work was performed in the frame of activities sponsored by the Italian National Group of Mathematical Physics (GNFM-INdAM) and by the Universities of Parma (Italy) and Pavia (Italy).
		RT has been supported by the National Institute of Advanced Mathematics (INdAM), Italy.	RT also thanks the support of the University of Parma through the action Bando di Ateneo 2022 per la ricerca, cofunded by MUR-Italian Ministry of Universities and Research - D.M. 737/2021 - PNR - PNRR - NextGenerationEU (project "Collective and Self-Organised Dynamics: Kinetic and Network Approaches").
		{The work of SB was supported by an appointment to the NASA Postdoctoral Program, held at NASA Goddard Space Flight Center and administered by Oak Ridge Associated Universities.}

		\appendix
		\section{Weakly nonlinear analysis in cross diffusion model}
		\label{appendix}
		\begin{itemize}
			\item Components of operator $\mathcal H_2\left[U_1, V_1
			\right]$ in  \eqref{EqOrd2.2Cr}
		\end{itemize}
		\begin{equation}
			\label{H_2_U}
			\begin{aligned}
				\mathcal H_2^U[U_1,V_1] =&\;
				\frac{c^3 V_1^2 (a \beta - c + 2 \zeta)}{\xi (c+\zeta)^2}
				+ \frac{2 a \beta c^3 \delta^c \, \Delta V_1 \, V_1}{\xi (c+\zeta)^2}
				+ \frac{2 a \beta c^3 \delta^c \nabla V_1\cdot\nabla V_1}{\xi (c+\zeta)^2} \\
				&- \frac{c^2 \zeta U_1 V_1 (4 a d + 1)}{\xi (c+\zeta)^2}
				- \frac{a c^2 D_2^1 \, \Delta V_1}{\xi (c+\zeta)}
				- \frac{c^2 \delta^c \, \Delta V_1 \, \zeta U_1}{\xi (c+\zeta)^2} \\
				&- \frac{2 c^2 \delta^c \zeta \, \nabla U_1\cdot\nabla V_1}{\xi (c+\zeta)^2}
				- \frac{c^2 \delta^c \, \Delta U_1 \, \zeta V_1}{\xi (c+\zeta)^2}
				- \frac{d \zeta^2 U_1^2}{(c+\zeta)^2}
				+ \frac{c D_2^1 \, \Delta U_1}{c+\zeta},
			\end{aligned}
		\end{equation}
		
		\begin{equation}
			\label{H_2_V}
			\begin{aligned}
				\mathcal H_2^V&[U_1,V_1] =\;
				\frac{4 a c^2 d \zeta U_1 V_1}{\xi (c+\zeta)^2}
				+ \frac{c^3 V_1^2 (c - 2 \zeta)}{\xi (c+\zeta)^2}
				+ \frac{d \zeta^2 U_1^2}{(c+\zeta)^2}.
			\end{aligned}
		\end{equation}

		\begin{itemize}
			\item Coefficients of the solution of \eqref{EqOrd2.2Cr} in the form \eqref{Expu2}
		\end{itemize}
		\begin{equation}
			\begin{aligned}
				X_0 =&\frac{
					2 c \left( 
					c^2 \left(-c^2 + c \zeta + \zeta^2 \right)
					- a c d \zeta (3c + \zeta) \rho
					- d \zeta^2 \xi \rho^2
					\right)
				}{
					a d \zeta (c + \zeta)^2 \xi
				}, \qquad
				Y_0 =\frac{
					2 \left(c^2 + a d \zeta \rho \right)^2
				}{
					c (c + \zeta)^2 \xi
				},\\[4mm]
				X_1=
				&-\Biggl[ 2 c^2 d \zeta \rho \Bigl( 
				a \left( 3 \delta^c k_c^2 - 1 \right) \left( \zeta (c + 3 D_Z k_c^2) + 3 c (c + D_Z k_c^2) \right) 
				- 12 D_Z k_c^2 \xi (c + \zeta) 
				\Bigr) \\
				&+ 2 c^3 \Bigl(
				c^2 \left( \zeta - 3 k_c^2 (a d D_Z + \delta^c \zeta) \right) 
				+ c \zeta \left( 3 k_c^2 \left( a d D_Z - \delta^c \zeta - 3 \delta^c D_Z k_c^2 + D_Z \right) + \zeta \right) \\
				&\quad + 3 D_Z \zeta^2 k_c^2 (2 a d - 3 \delta^c k_c^2 + 1) 
				+ c^3 (3 \delta^c k_c^2 - 1) 
				\Bigr) \\
				&- 2 d \zeta^2 \xi \rho^2 \left( c (3 a d D_Z k_c^2 - 3 c \delta^c k_c^2 + c) + 3 a d D_Z \zeta k_c^2 \right)
				\Biggr] \\
				&\times \Bigl[ d \xi (c + \zeta)^2 \Bigl(
				- a \zeta (c + 3 D_Z k_c^2) \left( 3 k_c^2 (D_2^c - \delta^c) + 1 \right) 
				+ 3 a c D_2^c k_c^2 (c - 3 D_Z k_c^2) 
				- 6 D_Z \zeta k_c^2 \xi
				\Bigr) \Bigr]^{-1},
			\end{aligned}
		\end{equation}
		\begin{equation}
			\begin{aligned}
				Y_1 = &\left[4 a c^2 d \zeta \rho \Bigl( 3 k_c^2 ( 2 c D_2^c + 2 D_2^c     \zeta  - \delta^c \zeta ) + \zeta \Bigr) \right.+ 2 c^3 \Bigl( 3 c^2 D_2^c k_c^2 + c \zeta (1 - 3 k_c^2 (D_2^c + \delta^c)) - 6 D_2^c \zeta^2 k_c^2 \Bigr) \\
				&\left.+ 2 d \zeta^2 \xi \rho^2 \Bigl( 3 k_c^2 (c D_2^c + \zeta (D_2^c - \delta^c)) + \zeta \Bigr)\right]\\
				&\times\left[a^2 d (c+\zeta)^2 \Bigl( 
				\zeta \Bigl( 3 D_Z k_c^2 (2 a d + 3 k_c^2 (D_2^c - \delta^c) + 1) + 3 c k_c^2 (D_2^c - \delta^c) + c \Bigr)\right. \\
				&\left.- 3 c D_2^c k_c^2 (c - 3 D_Z k_c^2) 
				\Bigr)\right]^{-1},
				\\[4mm]
				X_2 =
				&-\Biggl[
				c^2 d \zeta \rho \Bigl(
				a (4 \delta^c k_c^2 - 1) \left( 3 c^2 + c (4 D_Z k_c^2 + \zeta) + 4 D_Z \zeta k_c^2 \right)
				- 16 D_Z k_c^2 \xi (c + \zeta)
				\Bigr) \\
				&+ c^3 \Bigl(
				c^2 (\zeta - 4 k_c^2 (a d D_Z + \delta^c \zeta))
				+ c \zeta \left( 4 D_Z k_c^2 (a d - 4 \delta^c k_c^2 + 1) + \zeta - 4 \delta^c \zeta k_c^2 \right) \\
				&\quad + 4 D_Z \zeta^2 k_c^2 (2 a d - 4 \delta^c k_c^2 + 1)
				+ c^3 (4 \delta^c k_c^2 - 1)
				\Bigr) \\
				&- d \zeta^2 \xi \rho^2 \left( c (4 a d D_Z k_c^2 - 4 c \delta^c k_c^2 + c) + 4 a d D_Z \zeta k_c^2 \right)
				\Biggr] \\
				&\times \Bigl[
				d \xi (c + \zeta)^2 \Bigl(
				- a \zeta (c + 4 D_Z k_c^2) \left( 4 k_c^2 (D_2^c - \delta^c) + 1 \right)
				+ 4 a c D_2^c k_c^2 (c - 4 D_Z k_c^2)
				- 8 D_Z \zeta k_c^2 \xi
				\Bigr)
				\Bigr]^{-1},
				\\[4mm]
				Y_2 =&\left[2 a c^2 d \zeta \rho \Bigl( 4 k_c^2 (2 c D_2^c + 2 D_2^c \zeta   - \delta^c \zeta) + \zeta \Bigr)\right. + c^3 \Bigl( 4 c^2 D_2^c k_c^2 + c \zeta \left( 1 - 4 k_c^2 (D_2^c + \delta^c) \right) - 8 D_2^c \zeta^2 k_c^2 \Bigr) \\
				&\left.+ d \zeta^2 \xi \rho^2 \Bigl( 4 k_c^2 (c D_2^c + \zeta (D_2^c - \delta^c)) + \zeta \Bigr)\right]\\
				&\times \left[a^2 d (c + \zeta)^2 \Bigl( 
				\zeta \Bigl( 4 D_Z k_c^2 \left( 2 a d + 4 k_c^2 (D_2^c - \delta^c) + 1 \right) + 4 c k_c^2 (D_2^c - \delta^c) + c \Bigr) \right.\\
				&\left. - 4 c D_2^c k_c^2 (c - 4 D_Z k_c^2) 
				\Bigr)\right]^{-1}.
			\end{aligned}
		\end{equation}
		
		\begin{itemize}
			\item Components of operator $\mathcal H_3\left[U_1, V_1, U_2, V_2	\right]$ in  \eqref{EqOrd3.2Cr}
		\end{itemize}
		
		\begin{equation}
			\label{H_3_U}
			\begin{aligned}
				\mathcal H_3&^U(U_1,V_1,U_2,V_2) =\frac{\beta c^4 V_1^3 \,(-a \beta + 2c - 2\zeta)}{\xi (c+\zeta)^3}
				-\frac{3 a \beta^2 c^4 \delta^c \, \Delta V_1 \, V_1^2}{\xi (c+\zeta)^3}
				-\frac{6 a \beta^2 c^4 \delta^c \, V_1 \nabla V_1\cdot\nabla V_1}{\xi (c+\zeta)^3} \\
				&\quad
				+\frac{\beta c^3 U_1 V_1^2 (\zeta - 2 a d (c-2\zeta))}{\xi (c+\zeta)^3}
				+\frac{2 a \beta c^3 D_{2}^{1} \, \Delta V_1 \, V_1}{\xi (c+\zeta)^2}
				+\frac{2 a \beta c^3 D_{2}^{1} \nabla V_1\cdot\nabla V_1}{\xi (c+\zeta)^2} \\
				&\quad
				+\frac{2 a \beta c^3 \delta^c \, \Delta V_2 \, V_1}{\xi (c+\zeta)^2}
				+\frac{4 a \beta c^3 \delta^c \, \nabla V_1 \nabla V_2}{\xi (c+\zeta)^2}
				+\frac{2 c^3 V_1 V_2 (a \beta - c + 2\zeta)}{\xi (c+\zeta)^2} \\
				&\quad
				+\frac{2 a \beta c^3 \delta^c \, \Delta V_1 \, V_2}{\xi (c+\zeta)^2}
				-\frac{c^2 \zeta U_1 V_2 (4 a d + 1)}{\xi (c+\zeta)^2}
				-\frac{c^2 \zeta U_2 V_1 (4 a d + 1)}{\xi (c+\zeta)^2}-\frac{a c^2 D_{2}^{1} \, \Delta V_2}{\xi (c+\zeta)} \\
				&\quad
				+\frac{2 \beta c^3 \delta^c \zeta U_1 V_1 \, \Delta V_1}{\xi (c+\zeta)^3}
				+\frac{2 \beta c^3 \delta^c \zeta U_1 \nabla V_1\cdot\nabla V_1}{\xi (c+\zeta)^3}
				+\frac{4 \beta c^3 \delta^c \zeta V_1 \nabla U_1\cdot\nabla V_1}{\xi (c+\zeta)^3} \\
				&\quad
				+\frac{\beta c^3 \delta^c \zeta V_1^2 \, \Delta U_1}{\xi (c+\zeta)^3}
				-\frac{2 c^2 d \zeta^2 U_1^2 V_1}{\xi (c+\zeta)^3}
				-\frac{c^2 D_{2}^{1} \zeta U_1 \Delta V_1}{\xi (c+\zeta)^2}
				-\frac{a c^2 D_{2}^{2} \, \Delta V_1}{\xi (c+\zeta)}  \\
				&\quad
				-\frac{2 c^2 D_{2}^{1} \zeta \nabla U_1\cdot\nabla V_1}{\xi (c+\zeta)^2}
				-\frac{c^2 D_{2}^{1} \zeta V_1 \Delta U_1}{\xi (c+\zeta)^2}
				-\frac{c^2 \delta^c \zeta U_1 \Delta V_2}{\xi (c+\zeta)^2} \\
				&\quad
				-\frac{2 c^2 \delta^c \zeta \nabla U_1\cdot\nabla V_2}{\xi (c+\zeta)^2}
				-\frac{c^2 \delta^c \zeta U_2 \Delta V_1}{\xi (c+\zeta)^2}
				-\frac{2 c^2 \delta^c \zeta \nabla U_2\cdot\nabla V_1}{\xi (c+\zeta)^2} \\
				&\quad
				-\frac{c^2 \delta^c \zeta V_1 \Delta U_2}{\xi (c+\zeta)^2}
				-\frac{c^2 \delta^c \zeta V_2 \Delta U_1}{\xi (c+\zeta)^2}
				-\frac{2 d \zeta^2 U_1 U_2}{(c+\zeta)^2} 
				+\frac{c D_{2}^{1} \Delta U_2}{c+\zeta}
				+\frac{c D_{2}^{2} \Delta U_1}{c+\zeta}	,
			\end{aligned}
		\end{equation}
		
		\begin{equation}
			\label{H_3_V}
			\begin{aligned}
				\mathcal H_3&^V(U_1,V_1,U_2,V_2) =\;
				\frac{2 a \beta c^3 d \, U_1 V_1^2 (c-2 \zeta)}{\xi (c+\zeta)^3}
				+ \frac{4 a c^2 d \zeta \, U_1 V_2}{\xi (c+\zeta)^2}
				+ \frac{4 a c^2 d \zeta \, U_2 V_1}{\xi (c+\zeta)^2} \\
				&+ \frac{2 \beta c^4 V_1^3 (\zeta - c)}{\xi (c+\zeta)^3}
				+ \frac{2 c^3 V_1 V_2 (c-2\zeta)}{\xi (c+\zeta)^2}
				+ \frac{2 c^2 d \zeta^2 U_1^2 V_1}{\xi (c+\zeta)^3}
				+ \frac{2 d \zeta^2 U_1 U_2}{(c+\zeta)^2}.
			\end{aligned}
		\end{equation}
		\begin{itemize}
			\item Coefficients of system \eqref{SistAj_2}
			\begin{equation}\label{CoefSistAj_2}
				\begin{aligned}
					s_1 = & \frac{
						2 \tau \Bigl( 4 a c^2 d \zeta \rho + c^4 - 2 c^3 \zeta + d \zeta^2 \xi \rho^2 \Bigr)
						- 2 c^2 \left( \delta^c k_c^2 - 1 \right) (a \beta c - \zeta \rho)
					}{
						c D_2^c k_c^2 \rho^2 (c + \zeta)(a c - \xi \rho)
					},\quad
					\tilde s_1 = -\frac{2 \beta c}{\rho^2 (c + \zeta)},\\[4mm]
					s_2 = &
					\left\{
					a c^2 \Bigl[
					\beta c^2 \Bigl(-3 \beta + 6 d \rho \tau 
					+ \delta^c k_c^2 (3 \beta - 2 (Y_0 + 5 Y_2)) 
					+ 2 Y_0 + 2 Y_2 \Bigr)\right. \\
					& - 2 c \zeta \Bigl(
					6 \beta d \rho \tau 
					- 2 d \tau (X_0 + X_2 + \rho (Y_0 + Y_2)) 
					+ \beta \delta^c k_c^2 (Y_0 + 5 Y_2) 
					- \beta (Y_0 + Y_2) \Bigr) \\
					& + 4 d \zeta^2 \tau (X_0 + X_2 + \rho (Y_0 + Y_2))
					\Bigr] + 2 c^5 \tau (-3 \beta + Y_0 + Y_2)
					- 2 c^4 \zeta \tau (-3 \beta + Y_0 + Y_2) \\
					& + c^3 \zeta \Bigl(
					X_0 (\delta^c k_c^2 - 1) 
					+ X_2 (5 \delta^c k_c^2 - 1) 
					+ \rho (\delta^c k_c^2 - 1)(-3 \beta + Y_0 + Y_2) 
					- 4 \zeta \tau (Y_0 + Y_2) \Bigr) \\
					& + c^2 \zeta^2 \Bigl(
					6 d \rho^2 \tau 
					+ \delta^c k_c^2 (X_0 + 5 X_2 + \rho (Y_0 + Y_2)) 
					- X_0 - X_2 - \rho (Y_0 + Y_2) \Bigr) \\
					&\left. + 2 c d \zeta^2 \xi \rho \tau (X_0 + X_2)
					+ 2 d \zeta^3 \xi \rho \tau (X_0 + X_2)\right\}\left[
					c D_2^c k_c^2 \rho^3 (c+\zeta)^2 (a c - \xi \rho)\right]^{-1},
					\\[4mm]
					s_3=\;&\left\{
					2 a c^2 \Bigl[
					-\beta c^2 \Bigl(
					-3 \beta + 6 d \rho \tau 
					- \delta^c k_c^2 (-3 \beta + Y_0 + 4 Y_1) 
					+ Y_0 + Y_1 \Bigr)\right. \\
					& + c \zeta \Bigl(
					12 \beta d \rho \tau 
					- 2 d \tau (X_0 + X_1 + \rho (Y_0 + Y_1)) 
					+ \beta \delta^c k_c^2 (Y_0 + 4 Y_1) 
					- \beta (Y_0 + Y_1) \Bigr) \\
					& - 2 d \zeta^2 \tau (X_0 + X_1 + \rho (Y_0 + Y_1))
					\Bigr] - 2 c^5 \tau (-6 \beta + Y_0 + Y_1)
					+ 2 c^4 \zeta \tau (-6 \beta + Y_0 + Y_1) \\
					& - c^3 \zeta \Bigl(
					X_0 (\delta^c k_c^2 - 1) 
					+ X_1 (4 \delta^c k_c^2 - 1) 
					+ \rho (\delta^c k_c^2 - 1)(-6 \beta + Y_0 + Y_1) 
					- 4 \zeta \tau (Y_0 + Y_1) \Bigr) \\
					& - c^2 \zeta^2 \Bigl(
					12 d \rho^2 \tau 
					+ \delta^c k_c^2 (X_0 + 4 X_1 + \rho (Y_0 + Y_1)) 
					- X_0 - X_1 - \rho (Y_0 + Y_1) \Bigr) \\
					& \left.- 2 c d \zeta^2 \xi \rho \tau (X_0 + X_1)
					- 2 d \zeta^3 \xi \rho \tau (X_0 + X_1)\right\}\left[-c D_2^c k_c^2 \rho ^3 (c+\zeta )^2 (a c-\xi  \rho )\right]^{-1},
				\end{aligned}
			\end{equation}
			with $\tau=1-\sigma$.
		\end{itemize}
		
		\bibliographystyle{plain}
		\bibliography{biblio.bib}
		
	\end{document}